\documentclass[12pt,a4paper]{scrartcl}
\pdfoutput=1

\usepackage[utf8]{inputenc}
\usepackage[T1]{fontenc}
\usepackage{amsmath,amssymb,mathtools}
\usepackage[separate-uncertainty=true,retain-unity-mantissa=false]{siunitx}
\usepackage{slashed}
\usepackage{graphicx}
\usepackage{booktabs,multicol,multirow}
\usepackage[shortlabels]{enumitem}
\usepackage{hyperref}
\usepackage[capitalize]{cleveref}
\usepackage{xspace}
\usepackage[noblocks]{authblk}
\usepackage[dvipsnames]{xcolor}
\usepackage{soul}
\usepackage[textsize=tiny]{todonotes}
\usepackage{subcaption}
\usepackage{ulem}
\usepackage{bbold}
\usepackage{comment}

\usepackage[sorting=none,%
  citestyle=numeric-comp,%
  bibstyle=mynumeric,%
  giveninits=true]{biblatex}
\newcommand{\gev}{\;\text{GeV}\xspace}
\newcommand{\tev}{\;\text{TeV}\xspace}
\newcommand{\meter}{\;\text{m}\xspace}
\newcommand{\invfb}{\;\text{fb}^{-1}\xspace}
\newcommand{\pb}{\;\text{pb}\xspace}

\newcommand{\Oquqd}{\ensuremath{Q^{(1)}_{quqd}}\xspace}
\newcommand{\Oqule}{\ensuremath{\mathcal{O}^{(1)}_{lequ}}\xspace}
\newcommand{\OuP}{\ensuremath{Q_{u\Phi}}\xspace}
\newcommand{\OPq}{\ensuremath{Q^{(1)}_{\Phi q}}\xspace}
\newcommand{\OuG}{\ensuremath{Q_{uG}}\xspace}
\newcommand{\OuW}{\ensuremath{Q_{uW}}\xspace}
\newcommand{\OuB}{\ensuremath{Q_{uB}}\xspace}
\newcommand{\OSqD}{\ensuremath{Q_{SqD}}\xspace}
\newcommand{\OSuP}{\ensuremath{Q_{Su\Phi}}\xspace}
\newcommand{\OqdlN}{\ensuremath{Q_{qdlN}}\xspace}
\newcommand{\OqulN}{\ensuremath{Q_{qulN}}\xspace}
\newcommand{\OdueN}{\ensuremath{Q_{dueN}}\xspace}
\newcommand{\OqqNN}{\ensuremath{Q_{qqNN}}\xspace}
\newcommand{\OSSqD}{\ensuremath{Q_{SSDq}}\xspace}
\newcommand{\OSSuP}{\ensuremath{Q_{SSu\Phi}}\xspace}
\newcommand{\OuZp}{\ensuremath{Q_{uZ^\prime}}\xspace}

\newcommand{\order}[1]{\ensuremath{{\cal O}(#1)}}

\NewBibliographyString{refname}
\NewBibliographyString{refsname}
\DefineBibliographyStrings{english}{%
  refname = {Ref\adddot},
  refsname = {Refs\adddot}
}
\DeclareCiteCommand{\ccite}
  {%
  \ifnum\thecitetotal=1
    \bibstring{refname}%
  \else%
    \bibstring{refsname}%
  \fi%
  \addnbspace\bibopenbracket%
  \usebibmacro{cite:init}%
   \usebibmacro{prenote}}
  {\usebibmacro{citeindex}%
   \usebibmacro{cite:comp}}
  {}
  {\usebibmacro{cite:dump}%
   \usebibmacro{postnote}%
   \bibclosebracket}
\newrobustcmd*{\Ccite}{\bibsentence\ccite}

\addtokomafont{disposition}{\rmfamily}

\title{Topportunities at the LHC: \\ Rare Top Decays with Light Singlets}
\author[ ]{Henning Bahl\footnote{\href{mailto:hbahl@uchicago.edu}{hbahl@uchicago.edu}}}
\author[ ]{Seth Koren\footnote{\href{mailto:sethk@uchicago.edu}{sethk@uchicago.edu}}}
\author[ ]{Lian-Tao Wang\footnote{\href{mailto:liantaow@uchicago.edu}{liantaow@uchicago.edu}}}
\affil[ ]{Department of Physics and Enrico Fermi Institute, University of Chicago, 5720~South~Ellis~Avenue, Chicago, IL~60637~USA}
\date{}

\begin{document}
\maketitle

\begin{abstract}\noindent
    The discovery of the top quark, the most massive elementary particle yet known, has given us a distinct window into investigating the physics of the Standard Model and Beyond. With a plethora of top quarks to be produced in the High Luminosity era of the LHC, the exploration of its rare decays holds great promise in revealing potential new physics phenomena. We consider higher-dimensional operators contributing to flavour-changing-neutral-current top decays in the SMEFT and its extension by a light singlet species of spin 0, 1/2, or 1, and exhibit that the HL-LHC (and other future colliders) may observe many exotic top decays in a variety of channels. Light singlets which primarily talk to the SM through such a top interaction may also lead to distinctive long-lived particle signals. Searching for such long-lived particles in top-quark decays has the additional advantage that the SM decay of the other top quark in the same event provides a natural trigger.
\end{abstract}
\setcounter{footnote}{0}

\newpage

\tableofcontents


\section{Introduction}\label{sec:intro}
As we look forward to the continued success of the Large Hadron Collider project and its transformation into the High Luminosity era, we must adapt to the evolving strengths of the program. With the enormous number of proton collisions to be observed, we can search for new physics hiding in extremely rare processes---so long as their signatures are unique enough. 
In this work, we focus on the possibilities for discovering new physics associated with top quarks, as we anticipate the projected $\sim 10^{10}$ top quarks to be produced through Standard Model processes at the HL-LHC.

Having new physics appear in association with the top quark is a feature of familiar approaches to the hierarchy problem, as new degrees of freedom are introduced to cancel the top quark's contribution to the Higgs mass, and there is no small Yukawa to suppress their interactions with the SM fields. On the other hand, light BSM singlets feature in many of the most well-studied extensions of the SM, from axion-like particles to sterile neutrinos to dark photons. On these general grounds, we are motivated to search for all the possible ways by which light singlets may appear in association with top quarks, for which an effective field theory construction of operators is well-suited.

And indeed these expectations are borne out in many cases, with an illustrative example being the Froggatt-Nielsen approach to flavor physics \cite{Froggatt:1978nt}, in which a complex scalar $\phi$ has flavor-dependent couplings like $(\phi/M)^{n_{ij}} \tilde{\Phi} \bar Q_{Li} u_{Rj}$. The large third-generation Yukawas imply third-generation couplings at lower dimension and can result in the pseudo-Goldstone ALP modes coupling most readily to top quarks \cite{Carmona:2022jid}. 
Similarly, `flavorful' two-Higgs-doublet models (see e.g.\ \ccite{Das:1995df,Altmannshofer:2015esa,Altmannshofer:2016zrn}) lead naturally to rare top decay signatures recently studied in \ccite{Altmannshofer:2019ogm}.
And composite Higgs models \cite{Dimopoulos:1979es,Panico:2015jxa} may easily include singlet scalars \cite{Gripaios:2009pe,Banerjee:2018fsx} which inherit large top couplings.
Top-philic couplings can also arise naturally in models of flavored dark matter (e.g.\ \ccite{Kumar:2013hfa,Batell:2013zwa,Kilic:2015vka,Renner:2018fhh}), where the dark sector shares in our flavor symmetry and so inherits sizeable couplings to the top quark through the SM violation of the flavor symmetry by the Yukawas. 

In the present work, we pursue a general and bottom-up approach as part of the broader program of BSM top physics. While we focus our discussion on rare top decays, we note that single top quark production provides an additional complementary approach to probe new physics coupled to top quarks.\footnote{The question whether top decays in di-top quark production or single top quark production provides more sensitivity strongly depends on the specific new physics scenario and experimental final states. Example comparison studies can be found in \ccite{Agram:2013koa,Altmannshofer:2023bfk}.} We assume new physics particles mediating the top rare decay are heavy, and parameterize the interactions relevant for these processes by higher-dimensional operators.
We consider the operators with only SM particles as well as those extended by a single new, light singlet field of various spins, and systematically examine which higher-dimensional operators lead to top decays at observable rates. 

Existing experimental searches for rare top-quark decays focus on flavor-changing-neutral-current (FCNC) decays to SM final states. These include searches for top-quark decays into a light quark and a $Z$ boson~\cite{ATLAS:2012hfh,CMS:2012wao,CMS:2013knb,ATLAS:2015vhj,CMS:2017wcz,ATLAS:2018zsq}, a photon~\cite{ATLAS:2022per}, a Higgs boson~\cite{ATLAS:2015ncl,CMS:2016obj,ATLAS:2017tas,CMS:2017bhz,CMS:2021gfa,ATLAS:2018jqi,ATLAS:2018xxe,ATLAS:2022gzn}, or two leptons~\cite{CMS:2022ztx}. 
For other work on the prospects of rare top-quark decays to SM states, we refer to e.g.\ \ccite{Mahlon:1994us,Fox:2007in,Drobnak:2008br,Han:2013sea,Durieux:2014xla,Chala:2018agk,Balaji:2020qjg,Cremer:2023gne,Altmannshofer:2023bfk}.
In addition, also a few searches for top-quark decays into a light BSM particle have been performed. This includes searches for top quarks decaying into a charged Higgs boson (see e.g.\ the recent \ccite{ATLAS:2018gfm,CMS:2019bfg,CMS:2022jqc,ATLAS:2023bzb}) as well as a neutral scalar~\cite{ATLAS:2023mcc}. 
We mention also some recent phenomenological discussions of top decays which include light BSM (pseudo)scalars (e.g.\ \ccite{Carmona:2022jid,Banerjee:2018fsx,Bhattacharyya:2022umc}), fermions (e.g.\ \ccite{Alcaide:2019pnf}), or vectors (e.g.\ \ccite{Kong:2014jwa,Kim:2014ana}).
Notably, the direct constraints on the width of the top quark \cite{ATLAS:2017vgz} are uncertain to $\mathcal{O}(1)$ GeV. Hence, the search for exclusive final states remains the most sensitive probe of rare decays of top quarks.

We consider especially those cases in which the new light species may be long-lived, which can lead to distinct signals,
and we will examine the extent to which this possibility is consistent with their coupling to the SM in top-quark decay operators.
Whenever one adds new light states to the SM there are other constraints to be confronted, and we also discuss qualitatively the complementary constraints from low-energy probes of flavor physics.
The prospect of long-lived BSM particles has garnered increased attention in recent years \cite{Alimena:2019zri} due to their distinctive experimental signatures and resulting low backgrounds \cite{Narain:2022qud,Maltoni:2022bqs,Knapen:2022afb}, and will be an important component of collider searches moving forward \cite{Liu:2018wte,Liu:2020vur,Chiu:2021sgs,Blondel:2022qqo,Alipour-Fard:2018lsf,Alipour-Fard:2018rbc}.
In comparison with some of the commonly studied long-lived particle (LLP) production benchmarks, such as through the Higgs portal, the top quark to LLP processes studied here have a distinct advantage that the SM decay of the other top quark in the same event provides a natural trigger.

This article is organized as follows. In \cref{sec:ops}, we organize the pertinent top quark decay operators in the SMEFT extended by light singlets. In Section \ref{sec:topdecays} we discuss the top quark decay channels induced by these operators and give the rates. In Section \ref{sec:bsmdecays} we consider the possible signatures of the outgoing light singlets as well as the constraints from complementary flavor physics searches. We conclude in \cref{sec:conclusion}.


\section{Operators of interest}
\label{sec:ops}

To evince the worthwhile phenomenological opportunities present in studying FCNC top decays at HL-LHC, we study the different types of operators through which the top quark might decay. It is our interest to consider a representative array of operators leading to different signatures. Therefore, we include representative operators for every operator class but leave out operators for which we expect a similar collider phenomenology (for example in the class of four-fermion operators, we only include the scalar current operators but leave out the vector current operators). Then it is natural to orient ourselves within the SM effective field theory (SMEFT) (see e.g.~\ccite{Grzadkowski:2010es}) and organize our thinking around an expansion in powers of $\Lambda_\text{NP}^{-1}$, the scale of new physics. However, our EFT of the light fields will contain not only the SM but additional possible singlet fields of spin zero, half, and one. These are a new light scalar $S$, a right-handed neutrino $N$, or a vector $Z'^\mu$. Recently, \ccite{Song:2023jqm} has cataloged operators involving these new fields, so provides a useful check on our considerations.
Another possibility to introduce light BSM particles, which we do not consider here, is to add a second doublet to the SM Higgs sector (i.e., in the framework of Two-Higgs doublet models).

{\setlength{\tabcolsep}{0.6 em}
\renewcommand{\arraystretch}{1.3}
\begin{table}\centering
\large
\begin{tabular}{|c|c|} \hline 
\multicolumn{2}{|c|}{SM dim 6} \\ \hline
\Oquqd & $\left( \bar Q_{Li}^a u_{Rj} \right) \varepsilon_{ab} \left( \bar Q_{Lk}^b d_{Rl} \right)$  \\
\Oqule & $\left(\bar Q_{Li}^a u_{Rj} \right) \varepsilon_{ab} \left(\bar L_{Lk}^b e_{Rl} \right)$ \\
\OuP & $(\Phi^\dagger\Phi) (\bar Q_{Li} u_{Rj} \tilde\Phi)$ \\
\OPq & $(\Phi^\dagger i\overset{{}_{\leftrightarrow}}{D}_\mu \Phi) (\bar Q_{Li} \gamma^\mu Q_{Lj})$ \\
\OuG & $(\bar Q_{Li} \sigma^{\mu\nu} \mathcal{T}^A u_{Rj})\widetilde \Phi G_{\mu\nu}^A $ \\
\OuW & $(\bar Q_{Li} \sigma^{\mu\nu} u_{Rj}) \tau^I\widetilde \Phi W_{\mu\nu}^I $ \\
\OuB & $(\bar Q_{Li} \sigma^{\mu\nu} u_{Rj})\widetilde \Phi B_{\mu\nu} $ \\ \hline
\end{tabular}
\begin{tabular}{|c|c|}  \hline
\multicolumn{2}{|c|}{BSM dim 5} \\ \hline
\OSqD & $ S (\bar Q_{Li} \slashed{D} Q_{Lj})$ \\
\OSuP & $ S (\bar Q_{Li} u_{Rj} \widetilde\Phi)$ \\ \hline
\multicolumn{2}{|c|}{BSM dim 6} \\ \hline
\OqdlN & $\left( \bar Q_{Li}^a d_{Rj} \right) \varepsilon_{ab} \left(\bar L_{Lk}^b N\right)$  \\
\OqulN & $\left( \bar Q_{Li} u_{Rj} \right) \left(\bar N L_{Lk} \right)$  \\
\OdueN & $\left( \bar e_{Rj}^c u_{Rj} \right) \left(\bar d_{Rk} N\right)$  \\
\OqqNN & $\left( \bar Q_{Li} \gamma_\mu Q_{Lj} \right) \left(\bar N \gamma^\mu N\right)$  \\
\OSSqD & $S^2 (\bar Q_{Li} \slashed{D} Q_{Lj})$ \\
\OSSuP & $S^2 (\bar Q_{Li} u_{Rj} \widetilde\Phi)$ \\
\OuZp & $(\bar Q_{Li} \sigma^{\mu\nu} u_{Rj})\widetilde \Phi F^\prime_{\mu\nu} $\\ \hline
\end{tabular}
\caption{Overview of operators inducing top-quark decays into SM particles (left) and BSM particles (right). $Q_L$ is a Dirac spinor containing the left-handed quark doublet. The right-handed quarks are denoted by $u_R$ and $d_R$, respectively. $L_L$ and $e_R$ are the left- and right-handed lepton doublets/singlets, respectively. $i$, $j$, $k$, and $l$ are the quark and lepton generation indices. Other indices are suppressed. The Higgs doublet is called $\Phi$. $D$ denotes the SM covariant derivative, and $\slashed{D}\equiv \gamma^\mu D_\mu$. The SM vector-boson field strengths are called $G$, $W$, and $B$. The BSM singlets are called $S$ (scalar) and $N$ (Dirac fermion). The $Z^\prime$ field strength is denoted by $F^\prime$. The superscript ``$c$'' denotes charge conjugation. $\mathcal{T}^A = \lambda^A/2$, where $\lambda^A$ are the Gell-Mann matrices, $\tau^I$ are the Pauli matrices, and $\varepsilon_{ab}$ is the $SU(2)_L$ invariant tensor.}
\label{tab:opdim}
\end{table}}

We will concern ourselves with solely the lowest-dimensional operators up to dimension 6, which will suffice for us to find top quark decay channels involving either solely SM states or additionally any of our new light singlets. We list these operators in \cref{tab:opdim}, separating out those which involve solely SM particles and those which include a new light state. With only the SM, new operators through which the top quark might decay begin in SMEFT at dimension six. Restricting us to baryon- and lepton-number conserving operators (see e.g.\ \ccite{Andrea:2011ws,Dong:2011rh} for discussions of those type of operators), these operators can be divided into different categories: four-fermion operators, Yukawa-like operators, current interactions, and dipole operators. Our aim is to study a representative sample, rather than a comprehensive one, so we choose operators exhibiting an array of final state signatures. But there are many choices for possible tensor structures in the four-fermion operators especially, as well as another possibility for the Higgs-quark current-current interaction \OPq (see e.g.~\ccite{Grzadkowski:2010es}). 

We leave a detailed study comparing these possibilities to future work and here content ourselves to study examples of the possible final states. We have checked the full space of operators explicitly to confirm that the operators which we do not present results for give similar results (within one order of magnitude) for the branching ratios in the representative subspace for which we have chosen to present results. And so these offer similar discoverability in terms of the number of events possible to observe.

In the case where we add a BSM singlet scalar $S$, we can write down new operators at dimension five, which consist of a SM marginal operator with a scalar added on. There are multiple such operators containing the top quark, but they are related by the quark equations of motion. We non-redundantly choose one Yukawa-like and one kinetic-like operator in which the top decay amplitudes arise directly. At the level of dimension-4 operators, and ignoring flavor structure, the Standard Model quark sector has three kinetic operators and two Yukawa operators. Each of these could be multiplied by a scalar singlet to make a SMEFT operator which might induce an exotic top decay. However, they are related by three equations of motion (see e.g.~\ccite{Barzinji:2018xvu}) 
\begin{equation}
    i \slashed{D} Q \sim y_u^\dagger u \widetilde{\Phi} + y_d^\dagger d \Phi, \quad i \slashed{D} u \sim y_u q \tilde{\Phi}^\dagger, \quad i \slashed{D} d \sim y_d q \Phi^\dagger,
\end{equation}
which reduce the independent SMEFT operators from five to two. In different choices of bases, particular top decay channels may appear either from an insertion of a single higher-dimension vertex, or from a SMEFT vertex in concert with a marginal operator (e.g., the $t\to S W j$ originates directly from the $S Q\slashed{D}Q$ operator, whereas an additional SM operator insertion is needed if the $Qd \Phi$ operator is used.). The chosen basis will allow us to easily exhibit the different effects these operators may have in the phenomenology of the $S$ itself in \cref{sec:bsmdecays}. 

Similar operators appear at dimension six involving two scalar singlets, such that these operators do not destabilize $S$. The fermion singlet $N$ can enter into four-fermi operators like those in the SMEFT, and we consider examples involving one or two fermion singlets, but again do not comprehensively study all the possibilities. 

Turning to the BSM vector boson $Z'$, this can be used to construct an additional dimension-six dipole operator. We do not consider the case where the top quark is directly charged under a flavored $Z'$ gauge symmetry, which would allow new marginal operators contributing to exotic top decays e.g. $\bar u_{R3} \slashed{Z'} u_{R2}$. Since such gauge symmetries are not respected by the SM Yukawas, they must be broken far above the electroweak scale in order to generate the Yukawas, and then the $Z'$ will not be accessible in top decays. 
It would be interesting to understand precisely whether there is any room for observable $\bar t \to Z' \bar c$ at small gauge coupling where one could conceivably avoid the flavor constraints, but this requires more detailed flavor analysis than we pursue.

We will find branching ratios which result in thousands to millions of exotic top decays at the HL-LHC. So it is conceivable that even further suppressed decays could still result in many rare top decays with potentially visible signatures. Some amplitudes of top decays to SM final states through operators up to dimension 13 have recently been studied \cite{Bradshaw:2023wco}.
An interesting line of further pursuit would be to consider models introducing new light SM singlets where the leading effects on top quarks are in even higher-dimensional operators --- perhaps the case of an additional discrete symmetry requiring some number $n$ of light singlets as decay products. 
We leave such efforts on maximizing the rare top decay discovery potential to future work.


\section{Top-quark decay channels} \label{sec:topdecays}

Based on the operators presented in \cref{sec:ops}, we now discuss the various top-quark decay channels facilitated by these operators. We calculate the corresponding decay widths using \texttt{MadGraph5\_aMC@NLO} (version 3.4.2)~\cite{Alwall:2014hca,Alwall:2014bza} with the necessary \texttt{UFO} model file~\cite{Degrande:2011ua,Darme:2023jdn} generated using \texttt{FeynRules}~\cite{Christensen:2008py,Alloul:2013bka}. For calculating the expected number of events at the HL-LHC, we employ the SM prediction for the $t\bar t$ cross section at $13.6\tev$ of $923.6 \pb$\footnote{This value has been calculated at the next-to-next-to-leading order in perturbative QCD, including soft-gluon resummation at the next-to-next-to-leading logarithmic order, and using a top-quark mass of $172.5\gev$ (see \ccite{Czakon:2011xx} and references therein).} and a luminosity value of $3000 \invfb$. The numbers can easily be rescaled to estimate the number of events at future hadron colliders like the FCC-hh.\footnote{The $t\bar t$ cross section at a center-of-mass energy of 100~TeV is $\sim 35$~nb~\cite{Mangano:2016jyj}. The estimated total luminosity is $\sim 20\,\text{ab}^{-1}$.}

\begin{figure}
  \centering
  \includegraphics[width=\textwidth]{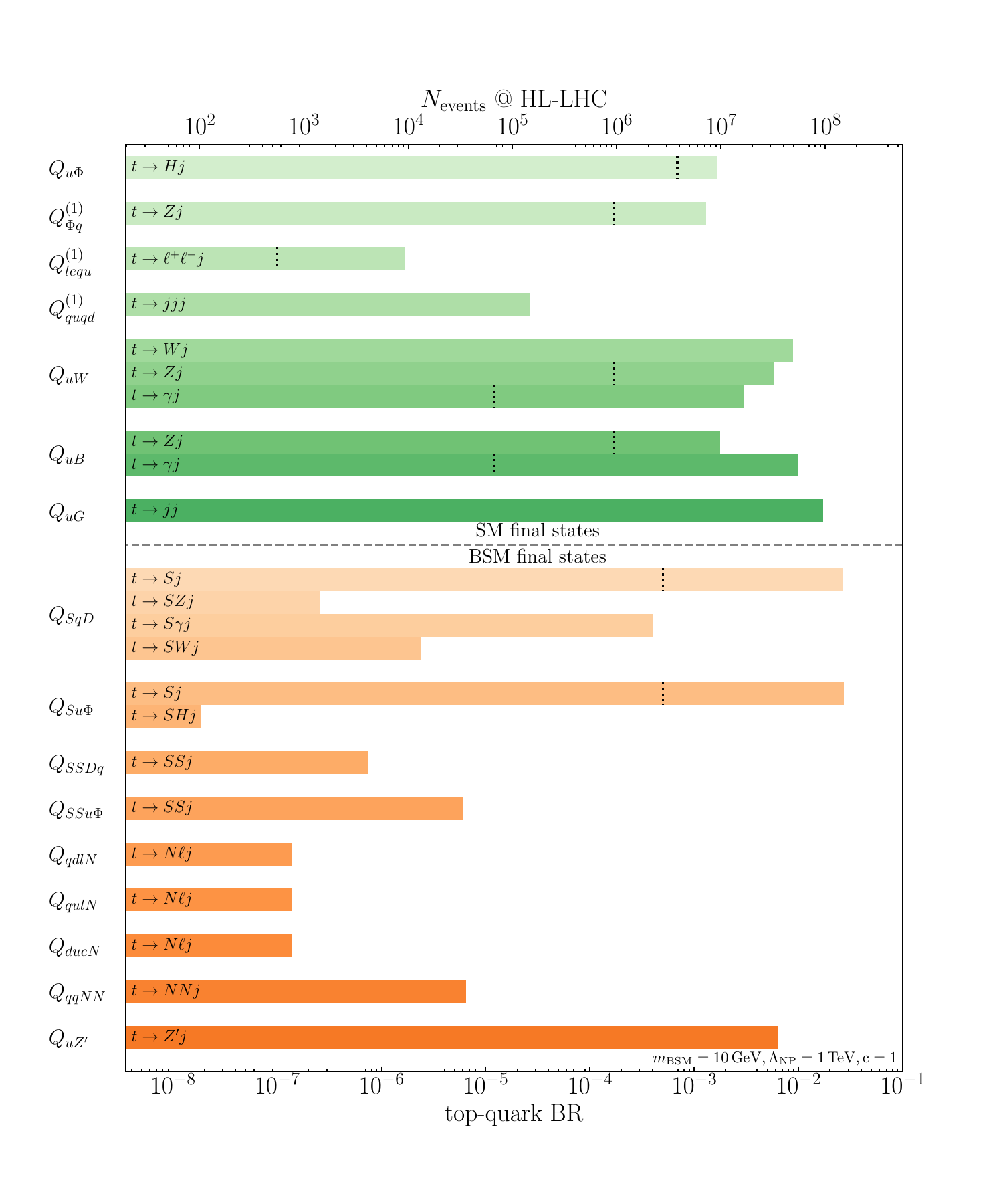}
  \caption{Branching ratios (lower axis) and expected number of events at HL-LHC (upper axis) for the various top-quark decay channels induced by operators listed on the left-hand side. The results for SM final states are shown in green colors at the upper side of the plot; the results for BSM final states, in red colors at the bottom. The numerical results have been derived setting $\Lambda_\text{NP} = 1\tev$ and all Wilson coefficients equal to one. Moreover, all masses of BSM particles have been set to $10\gev$. The black dotted lines indicate existing collider constraints.}
  \label{fig:summary}
\end{figure}

An overview of the branching ratios for the various decay channels is presented in \cref{fig:summary}. For this Figure, we set $\Lambda_\text{NP} = 1\tev$. All Wilson coefficients are fixed to one ($c = 1$). Regarding the flavour structure of the operators, we choose to set the coupling involving the top and the first-generation quarks to a non-zero value, while the coupling involving the second-generation quarks is fixed to zero. Consequently, the quark in the final state is a first-generation quark (similar results are expected for second-generation quarks in the final state). For final state photons, we place a lower transverse momentum cut of $20\gev$ to avoid infrared divergences.

In the upper part of \cref{fig:summary}, the branching ratios for the top-quark decays into SM final states are shown.\footnote{Additional higher-multiplicity decay modes are possible but not shown due to a highly suppressed rate.} The four-fermion operators induce decays either into three jets or into two leptons and a jet. The corresponding branching ratios are of \order{10^{-5}} corresponding to \order{10^4} events expected at HL-LHC. The various decay channels induced by dipole operators ($t\to Wj, Zj, \gamma j, jj$) generally yield higher branching ratios of \order{10^{-2}} corresponding to \order{10^7-10^8} events expected at HL-LHC. Most of these decay channels have already been probed experimentally. The currently strongest experimental 95\% confidence level upper limits are
\begin{itemize}
    \item $\text{BR}(t\to Z u) < 1.7\cdot 10^{-4}$ and $\text{BR}(t\to Z c) < 2.4\cdot 10^{-4}$~\cite{ATLAS:2018zsq},
    \item $\text{BR}(t\to \gamma u) < 1.2\cdot 10^{-5}$ and $\text{BR}(t\to \gamma c) < 4.5\cdot 10^{-5}$~\cite{ATLAS:2022per},
    \item $\text{BR}(t\to H u) < 6.9\cdot 10^{-4}$ and $\text{BR}(t\to H c) < 9.4\cdot 10^{-4}$~\cite{ATLAS:2022gzn},
    \item $\text{BR}(t\to u e \mu) \lesssim 10^{-7}$ and $\text{BR}(t\to c e \mu) < 10^{-6}$~\cite{CMS:2022ztx}.
\end{itemize}
These limits are indicated in \cref{fig:summary} in the form of dotted black lines.
There may additionally be indirect, low-energy constraints on these SMEFT operators from flavor physics observables. We refrain from placing these in the above figure, as these complementary data do not diminish the usefulness of direct searches, but we refer to recent works with which such limits may be derived e.g.~\ccite{DeBlas:2019ehy,Aebischer:2018iyb,Straub:2018kue,GAMBITFlavourWorkgroup:2017dbx,EOSAuthors:2021xpv}.

In the lower part of \cref{fig:summary}, the branching ratios for the top-quark decays into one or more BSM particles (plus accompanying SM particle(s)) are shown setting the BSM masses to $10\gev$. As is visible, large branching ratios of \order{10^{-2} - 10^{-1}} can be obtained for two-body final states containing either $S$ or $Z^\prime$. The decays into a light scalar and a jet are already constrained by a recent experimental search which constrains $\text{BR}(t\to u S,c S)\cdot\text{BR}(S\to b \bar b) \lesssim 10^{-4}-10^{-3}$ (for $20\gev < m_S < 160\gev$)~\cite{ATLAS:2023mcc}. For depicting this limit in the \cref{fig:summary}, we assume a $\text{BR}(S\to b\bar b) = 1$ and extrapolate the limit from $m_S = 20\gev$ to $10\gev$. This search can, however, be completely evaded if $S$ decays less-frequently to bottom quarks. The operators \OSqD and \OSuP also induce further three-body decays with an additional gauge boson in the final state. Their branching ratios are significantly smaller making them phenomenologically less relevant (with the potential exception of the $t\to S\gamma j$ channel, where the additional photon could help with background separation). One additional final state with one BSM particle is the $t\to N\ell j$ decay. This channel has a comparably low branching ratio of \order{10^{-6}}, which would, however, still result in a sizeable number of events at the HL-LHC. The branching ratios with two BSM particles in the final state are of similar size. Apart from the decay channels shown in \cref{fig:summary}, a few other possibilities like $t\to Z^\prime H j$ or $t \to S S Z j$ exist which are, however, highly suppressed.

We also want to stress here again that the provided numbers are only a rough indication of the potential size of these decay modes. The results are derived for a fixed UV scale $\Lambda_\text{NP}$ (but can easily be rescaled). This scale (and also the Wilson coefficients) could very well be lower or higher.

\begin{figure}
  \centering
  \includegraphics[width=.49\textwidth]{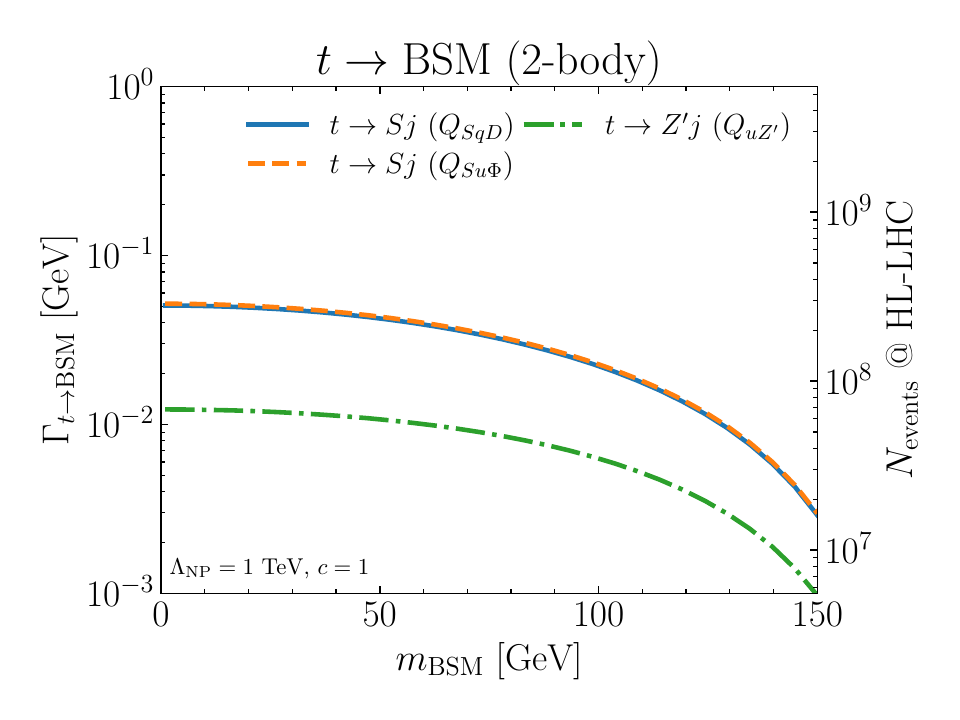}
  \includegraphics[width=.49\textwidth]{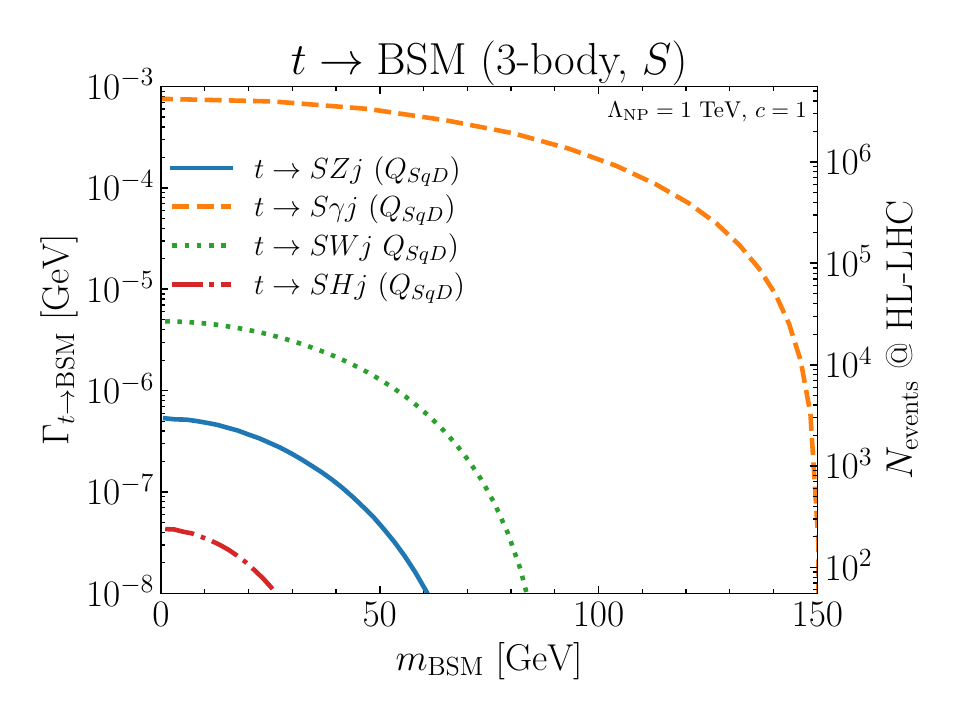}
  \includegraphics[width=.49\textwidth]{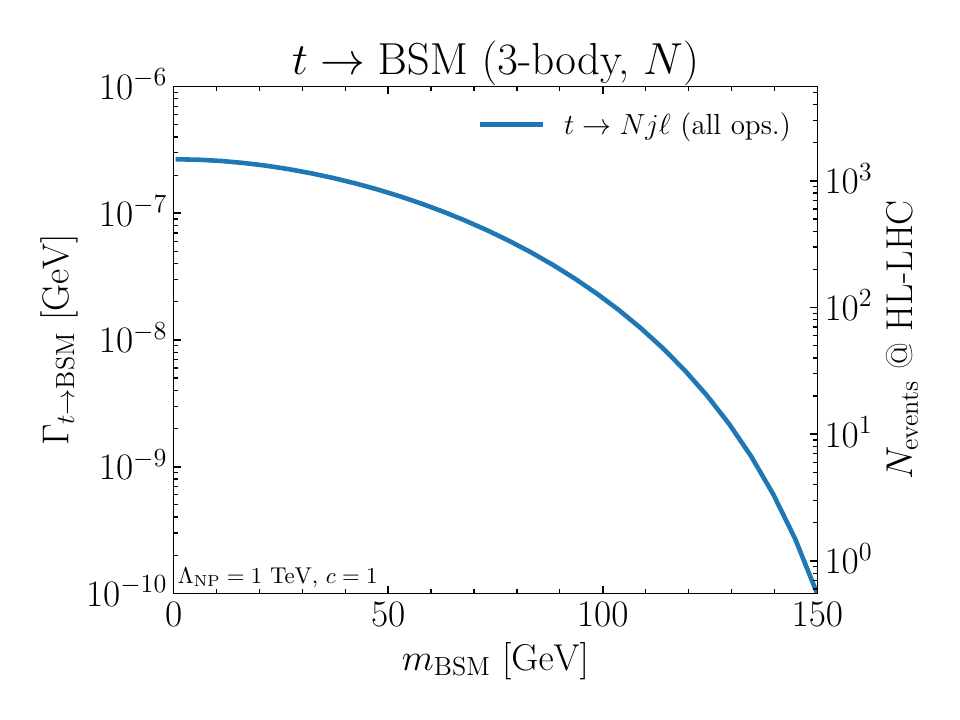}
  \includegraphics[width=.49\textwidth]{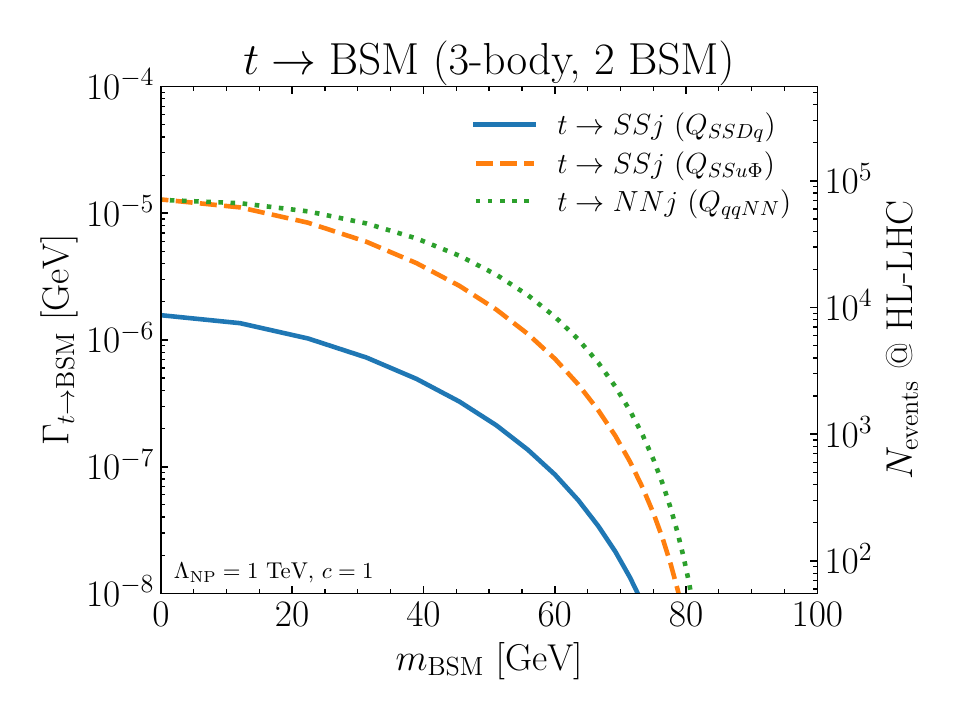}
  \caption{Decay widths for top-quark decays including BSM final states as a function of the mass of the BSM particle. The results are shown setting $\Lambda_\text{NP} = 1\tev$ and all Wilson coefficients equal to one.}
  \label{fig:t_to_BSM}
\end{figure}

We study the mass dependence of the top quark decays into one or two BSM particles in \cref{fig:t_to_BSM}. The two-body decay widths are shown in the upper left panel. As expected, the decay widths shrink if the mass of the BSM particle in the final state is raised (by up to an order of magnitude in the considered mass window). The dependence on the BSM mass is stronger for higher masses, where the phase space is rapidly shrinking.

The same behaviour is visible for the three-body final states with one BSM particle in the final state (see upper right and lower left panels of \cref{fig:t_to_BSM}). For final states with two BSM particles the drop of decay widths is even more rapid (see lower right panel of \cref{fig:t_to_BSM}).

As mentioned above, an additional possibility for rare top-quark decays can be introduced by adding an additional scalar doublet to the SM Higgs sector. Such a Two-Higgs-Doublet model can lead to top-quark decays into a bottom quark as well as a charged Higgs boson $H^\pm$. Since this possibility has been widely studied in the literature --- discussing cases in which the charged Higgs boson decays directly to SM particles or first into other BSM states (see e.g.~\ccite{Akeroyd:2016ymd,Bahl:2021str,Cheung:2022ndq,Hu:2022gwd,Krab:2022lih,Bhatia:2022ugu,Kim:2023lxc,Mondal:2023wib,Fu:2023sng,Sanyal:2023pfs,Li:2023btx} for a review and recent works) --- as well as probed experimentally (see e.g.\ the recent \ccite{ATLAS:2018gfm,CMS:2019bfg,CMS:2022jqc,ATLAS:2023bzb}), we do not consider it in this work.

In summary, \cref{fig:summary} provides a good starting point (and strong motivation) for future experimental searches. The reach of these searches will depend strongly on the specific final state. As we will show below, for many of the discussed channels, the BSM particles in the final state can be long-lived providing a unique signal. We expect that such signals, which are mostly background-free, provide a large reach (especially given that the second top quark provides a natural trigger candidate). We leave detailed studies for these long-lived particle signatures, which require detailed modelling of the detector, for future work.

In the more-difficult case where the decay is prompt, we briefly comment on the expected backgrounds at HL-LHC. We first consider the case with one BSM particle in the final state. The difficulty in distinguishing these rare decays from SM backgrounds depends sensitively on the decay modes of the BSM particle. If the singlet decay includes higher multiplicities of $b$-jets or leptons, such as $S, Z' \rightarrow b\bar b (\ell \bar \ell)$ or $N \rightarrow b\bar b \nu (\ell \bar \ell \nu)$, this provides a handle to reduce the SM $t\bar t$ background. The additional backgrounds are $t\bar t + W,Z,h$, $t\bar t b \bar b$ production (if the singlet decays to $b$ quarks), $t\bar t \gamma\gamma$ production (if the singlet decays to photons), whose cross-sections are down by a factor of $\sim 10^3$~\cite{Broggio:2019ewu} in comparison to the $t\bar t$ cross-section with $\sim 10^6$ such events expected at HL-LHC. As projected in \ccite{Banerjee:2018fsx} a careful analysis employing multivariate analysis techniques allows setting a limit on the rare top branching ratio of up to for the $t\to S j$ with $S\to b\bar b$ ($S\to\gamma\gamma$) sensitivity on the branching ratio reaching down to $10^{-4}$ ($10^{-7}$). For the $S\to b\bar b$, this potential has been explicitly confirmed by the ATLAS collaboration~\cite{ATLAS:2023mcc}. Moreover, possible leptonic decay modes of the BSM singlet have been studied in detail in \ccite{Bhattacharyya:2022ciw}. On the other hand, decay modes to light quarks will be more difficult to distinguish by particle identification alone.

Top decay modes with two BSM particles are expected to have a lower rate. They, however, also profit from lower backgrounds due to the larger number of final state particles. In this case, the question of which backgrounds are most relevant will even more sensitively depend on the concrete decay modes. While decays into light quarks will again be very difficult to distinguish from SM multi-jet backgrounds, leptonic/bottom decay modes will be easier with the most dominant background being multi-top plus jets/vector-boson production.

While prompt decays are more challenging than the long-lived modes we focus on below, \cref{fig:summary} still provides a starting point for further analysis of search strategies for the most promising rare top decay modes.


\section{Light BSM singlet particles} \label{sec:bsmdecays}

In this Section, we discuss the phenomenology of the light singlets produced in FCNC rare top decays.


\subsection{Signals from the decays of BSM singlet particles}

The phenomenology of light singlet particles produced at colliders will depend sensitively on their lifetimes $\tau$. There are three different possibilities:
\begin{itemize}
    \item \textbf{Prompt decaying BSM particles ---} Prompt decays into SM particles, with a decay length $c \tau \lesssim 1 \,\text{mm}$, can be induced by introducing additional operators with a large enough Wilson coefficient. While the inclusion of multiple operators involving the new light singlet fields is in some sense `non-minimal', one realistically expects that many operators are generated in a UV completion barring some additional structure forbidding them. For example, the $N$ may interact via the relevant `neutrino portal' $y_N \Phi L_L N$ leading to mixing with the SM neutrinos. The $S$ and $Z^\prime$ singlets could e.g.\ decay into bottom quarks, light quarks, leptons, or two photons. This case has been studied in detail in \ccite{Banerjee:2018fsx,Bhattacharyya:2022ciw,Bhattacharyya:2022umc}.

    An additional, irreducible possibility for prompt decays into SM particles is a decay involving the operator(s) responsible for the top-quark decay into the light BSM particle. This case is studied in detail below.

    \item \textbf{Stable BSM particles ---} For the top-quark decay operators involving two BSM particles, the respective BSM particle is stable if no additional operators involving only one instance of the BSM field are introduced. From a UV perspective, such a situation can be generated e.g.\ by imposing a $\mathbb{Z}_2$ symmetry. In this case, the BSM particles would manifest themselves in the form of a missing transverse energy signal.

    \item \textbf{Long-lived BSM particles ---} The BSM particles can also be long-lived ($1 \,\text{mm} \lesssim c \tau$). The most striking signatures possible appear when the decays are displaced from the interaction point and result in a rich phenomenology. Such signatures typically have no SM backgrounds, and so the threshold for discovery may be just a few events at the HL-LHC, making extremely rare such processes nonetheless visible. One such signature has been studied in \ccite{Carmona:2022jid}, investigating the decay of top quarks into long-lived axion-like particles.

    While this is the most interesting possibility, we are not necessarily free to dictate it, as there may be an upper bound on the lifetime of the new state we are considering induced by the top-quark decay operator through which it was produced. We will calculate such upper bounds below. Additional decay channels can of course always be opened by introducing additional operators with a small enough Wilson coefficient (or a large enough scale suppression).

    Although detector sizes are typically $\lesssim 10 \meter$, for larger lifetimes the rate of decay in the detector falls only linearly in $c\tau$ and it is proportional to the volume of the detector component which is sensitive to the signal, allowing ample detection prospects on the tail of the decay distribution. The search for LLPs in rare top-quark decays has a distinct advantage in comparison to some of the other production mechanisms. For example, a well-studied benchmark is the production of LLPs from rare Higgs-boson decay through the Higgs portal coupling. Typically, additional hard radiation is needed to trigger on this kind of signal. However, in the case of rare top-quark decay, the SM decay of the other top in the same event provides a natural trigger without further suppressing the signal rate. Eventually, these modes become too rare, and the signature in our colliders is simply missing transverse energy mimicking the situation in which they are stable. This limits the information one may uncover about the new states even if such exotic top decays are observed.
\end{itemize}

\begin{figure}
  \centering
  \includegraphics[width=.28\textwidth]{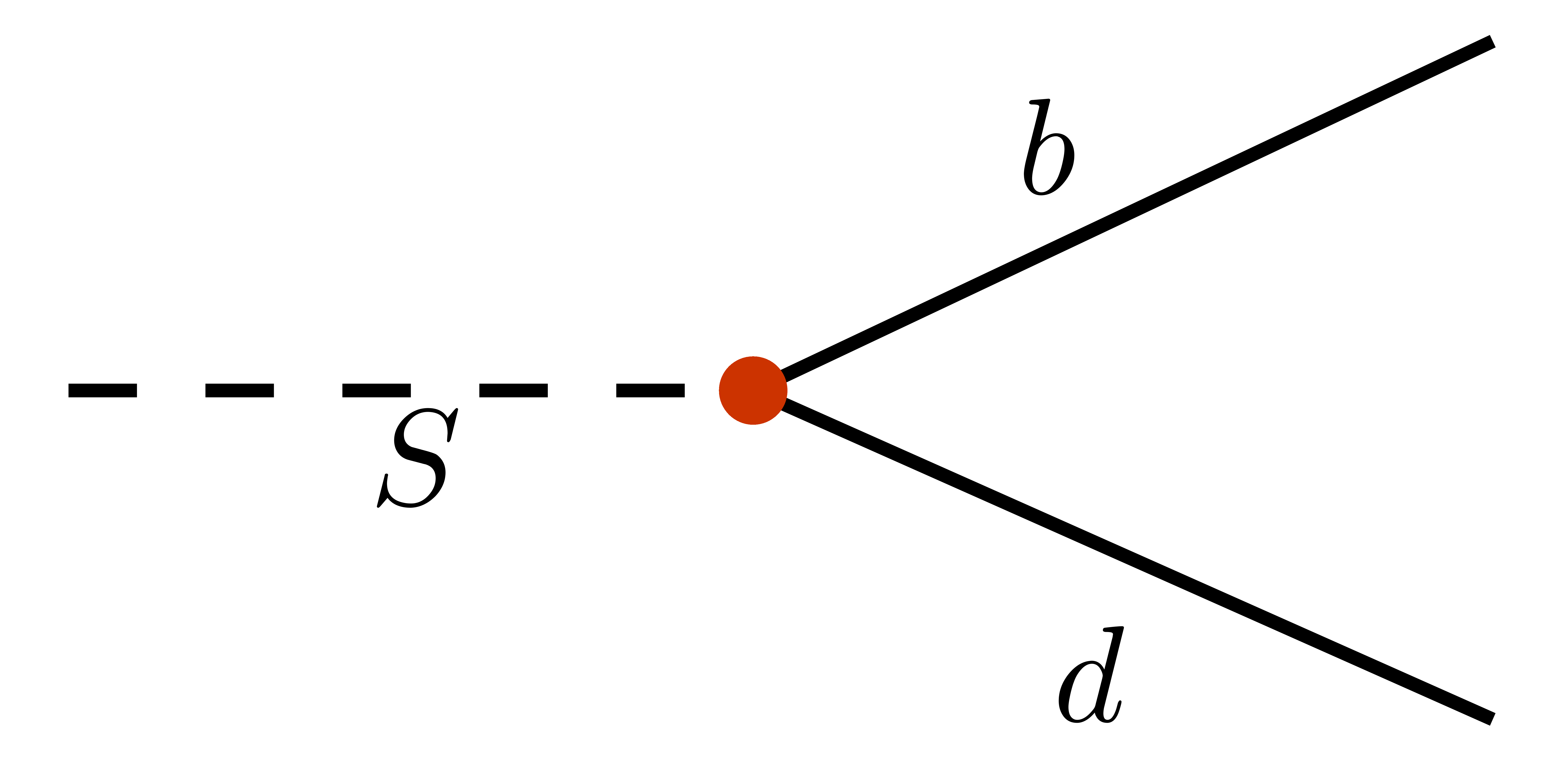}\hspace{.5cm}
  \includegraphics[width=.32\textwidth]{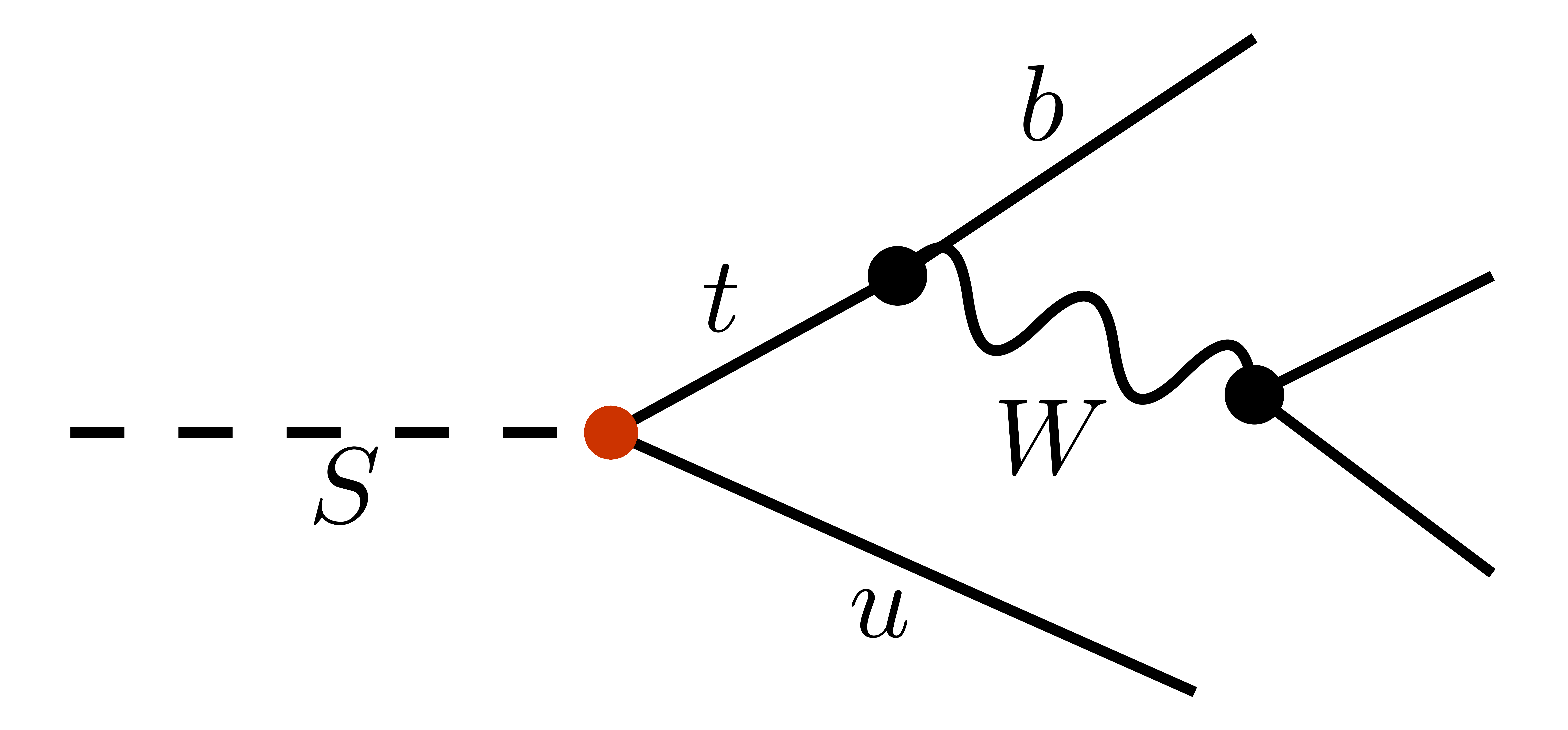}\hspace{.5cm}
    \includegraphics[width=.32\textwidth]{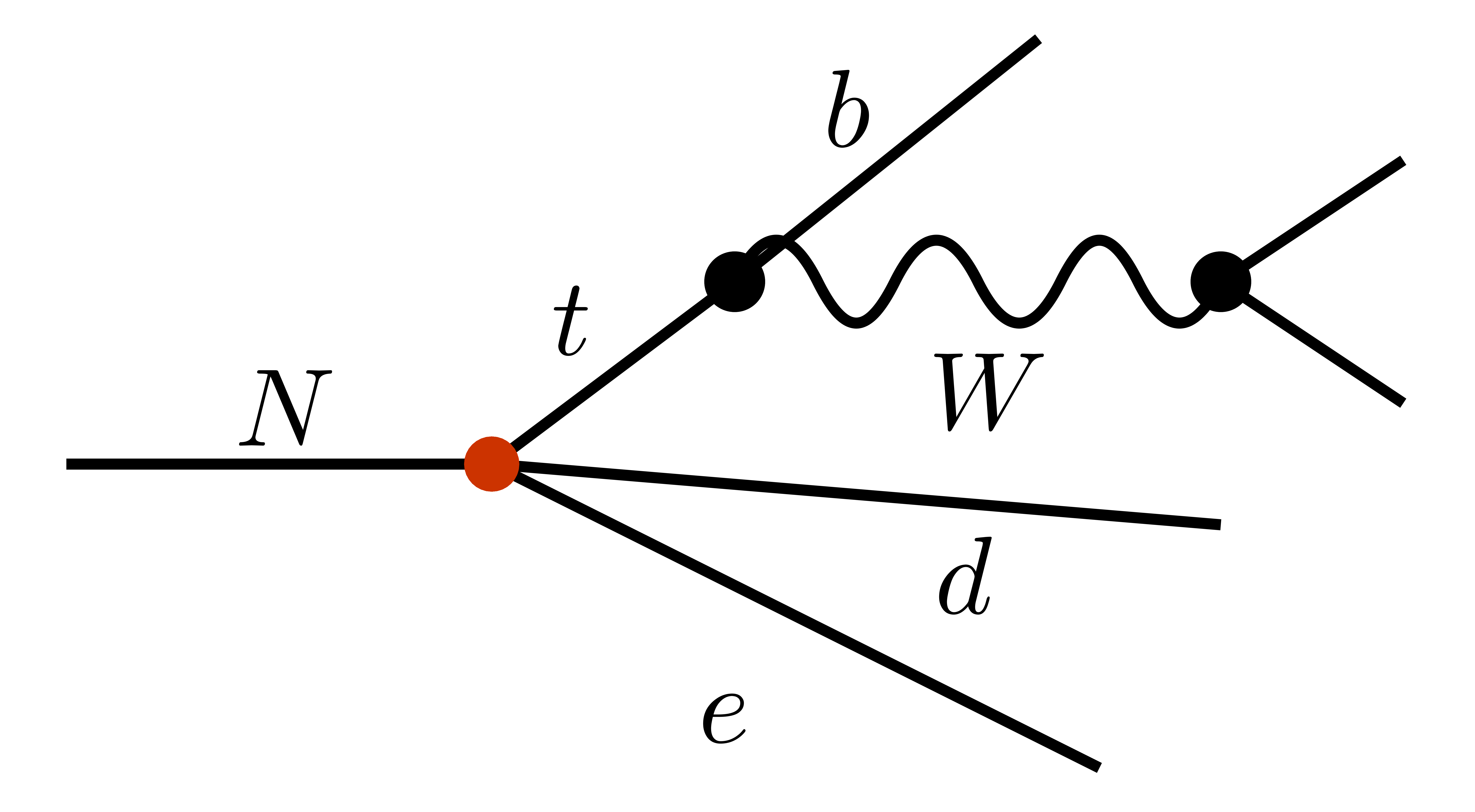}\hspace{.5cm}
  \caption{Feynman diagrams for tree-level decays of the scalar singlet $S$ when the new physics interaction involves the $b$-quarks directly (left panel) or when an off-shell $W$ appears in the intermediate diagram (middle panel). Analogous tree-level decays can occur for $Z^\prime$. In the right panel, we show a five-body tree-level decay of $N$ through an off-shell $W$. The red vertex denotes the dimension five or six operators responsible for the respective top-quark decay into a light BSM particle.}
  \label{fig:tree_Sdecays_Feynman}
\end{figure}

\begin{figure}
  \centering
  \includegraphics[width=.28\textwidth]{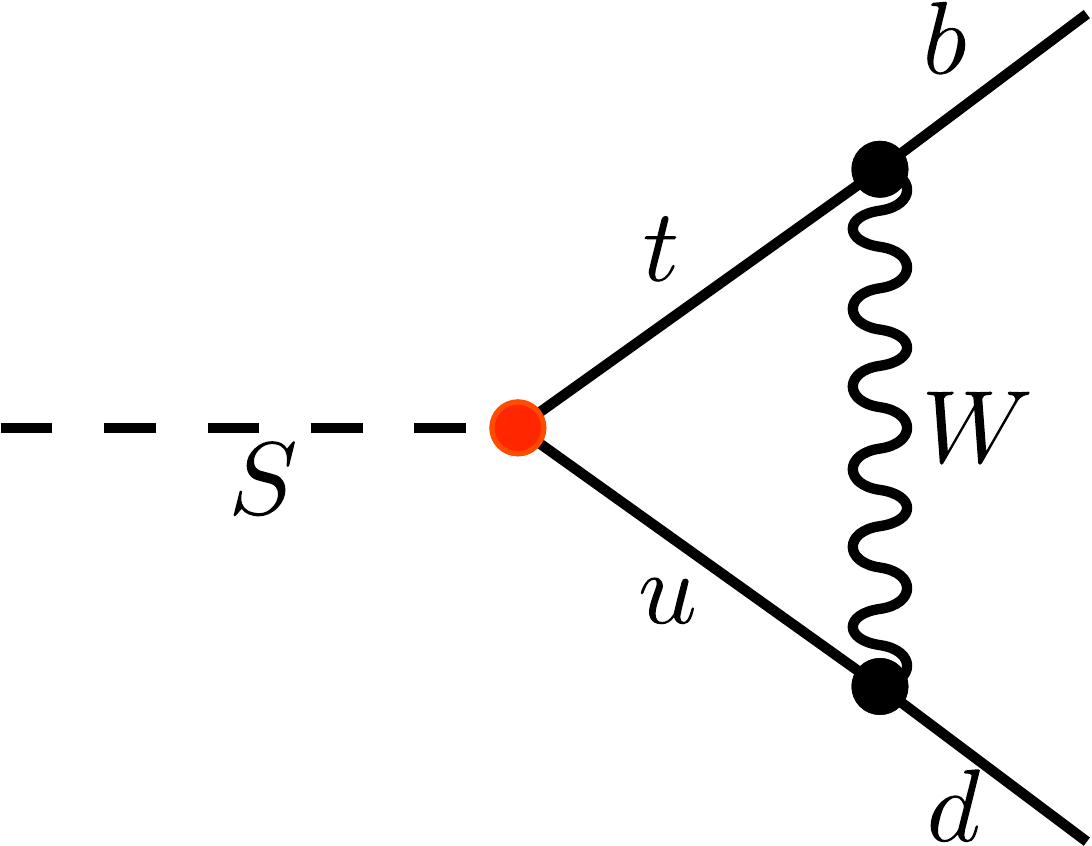}\hspace{.5cm}
  \includegraphics[width=.28\textwidth]{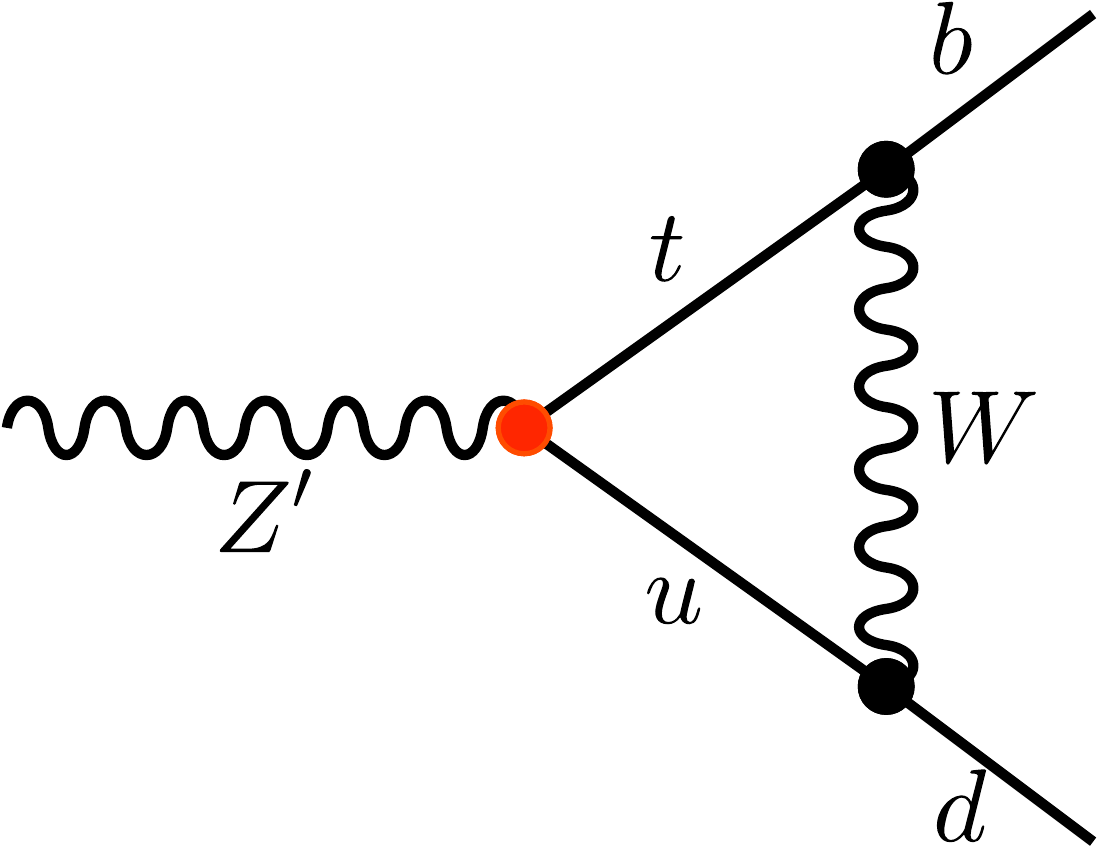}\hspace{.5cm}
  \includegraphics[width=.28\textwidth]{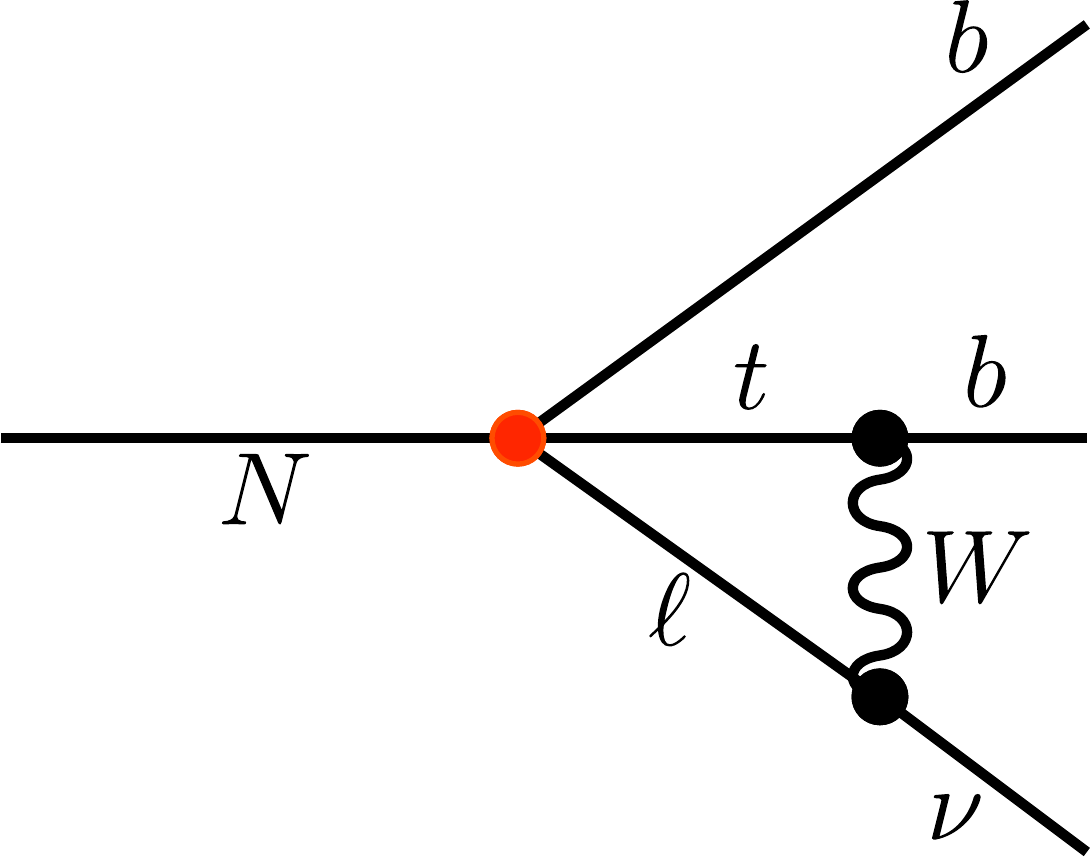}
  \caption{Feynman diagrams for loop-induced decays of the scalar singlet $S$ (left panel), the gauge boson $Z^\prime$ (middle panel), and the sterile neutrino $N$ (right panel). The red vertex denotes the dimension five or six operator responsible for the respective top-quark decay into a light BSM particle.}
  \label{fig:lightBSM_decays_Feynman}
\end{figure}

We proceed with a calculation of the lower bounds on the decay width of the BSM particles originating from diagrams involving the top-quark decay operators. We consider three different possibilities:
\begin{itemize}
    \item The decay is induced at the tree level involving no off-shell particles. Such a decay occurs for example for the \OSqD operator, which allows for $S \to b d$ or $S \to b s$ as the operator is constructed from the left-handed quark doublets (see left diagram of \cref{fig:tree_Sdecays_Feynman}).
    \item The decay is induced at the loop level with an internal $W$ boson to exchange the top with a lighter quark, as in \cref{fig:lightBSM_decays_Feynman}.
    \item The decay is induced at the tree level involving an off-shell top-quark and a (potentially off-shell) $W$ boson, which further decays either hadronically or leptonically (see middle and right diagrams of \cref{fig:tree_Sdecays_Feynman}).
\end{itemize}
To calculate these decay rates, we employ \texttt{MadGraph5\_aMC@NLO} for the tree-level decays and the packages \texttt{FeynArts}~\cite{Kublbeck:1990xc,Hahn:2000kx}, \texttt{FormCalc}~\cite{Hahn:1998yk}, and \texttt{LoopTools}~\cite{Hahn:1998yk} for the loop-induced decays---having generated the necessary model file using \texttt{FeynRules}. 

Note that the appearance of a $W$ boson in the later two possibilities to exchange the heavy top for the lighter bottom means that these decay widths can be sensitive to the interplay of flavor with chirality in the operators. For example, for an operator containing $\bar Q_{L1} u_{R3}$, the right-handed top quark can be easily switched to left-handed by its large Yukawa, such that a $W$ boson can easily be coupled to the top quark resulting in sizeable loop-induced decay rates. But in the opposite case $\bar Q_{L3} u_{R1}$ it is very costly to change the up-quark chirality, and the leading decay channel may become the four-body tree-level decay with an off-shell $W$. These considerations made little difference for the top-quark decay rates but can have a large impact on the phenomenology of the light singlets.

\begin{figure}
  \centering
  \includegraphics[width=.48\textwidth]{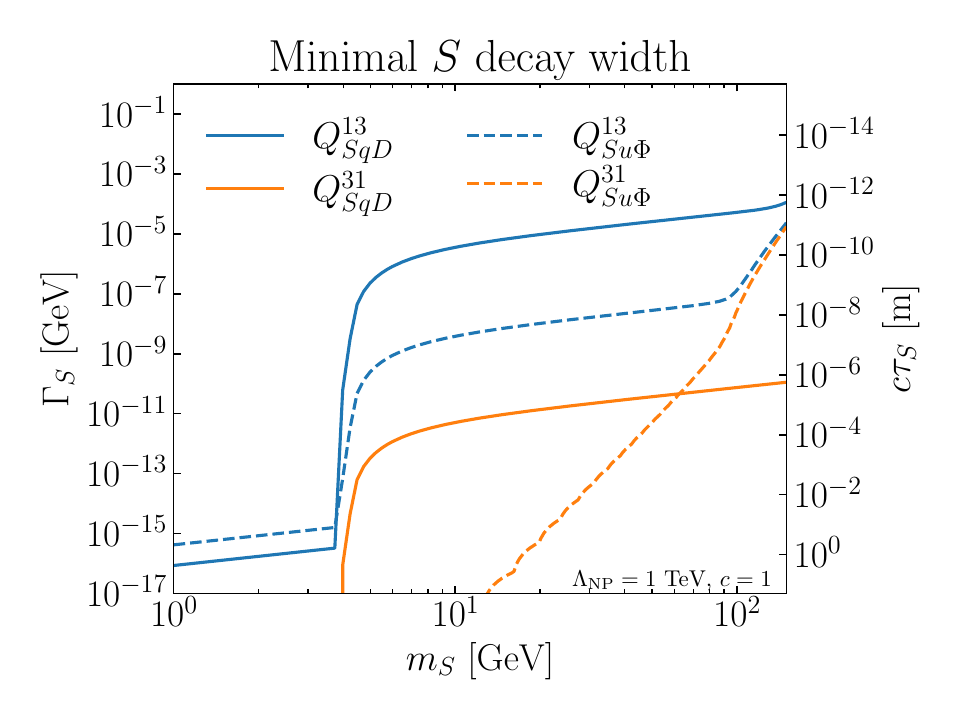}
  \includegraphics[width=.48\textwidth]{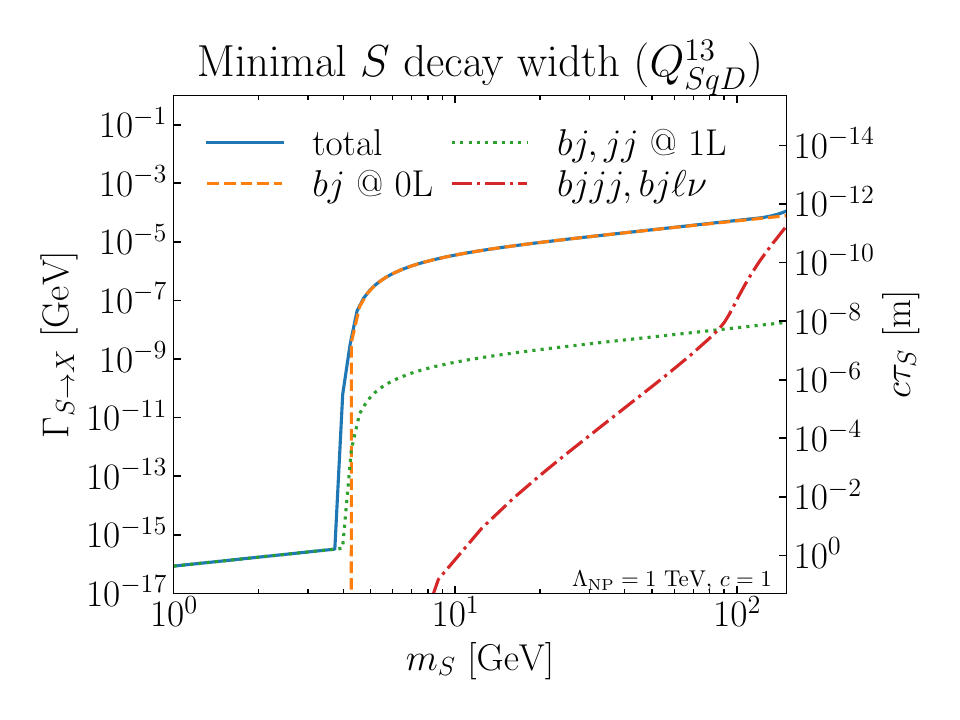}
  \includegraphics[width=.48\textwidth]{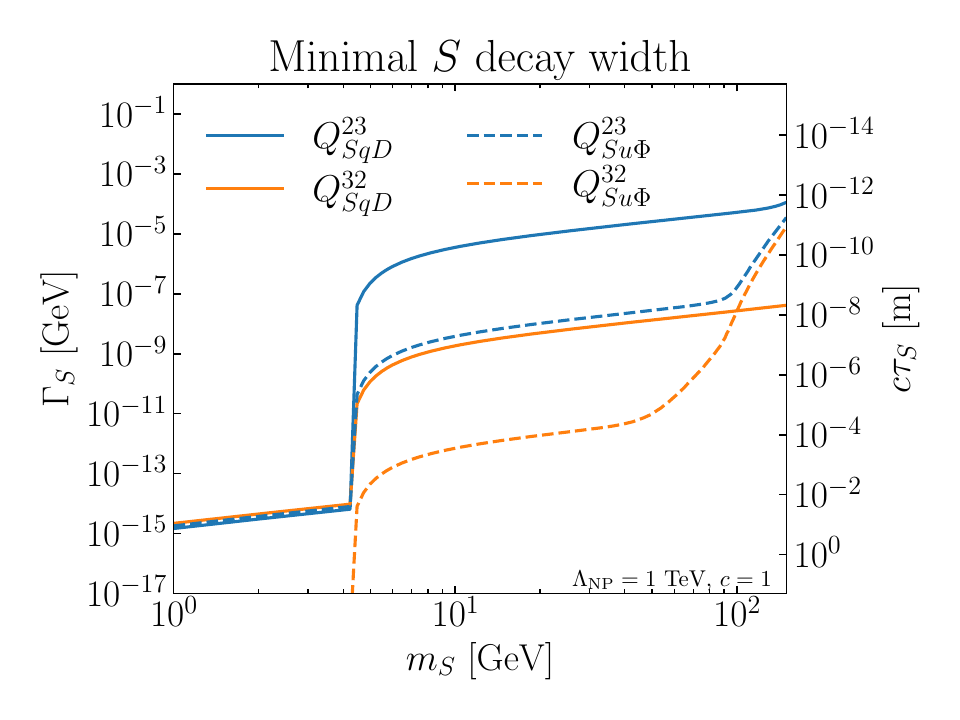}
  \caption{Upper left: Mass dependence of the decay width of the scalar singlet $S$ for operators coupling the top quark to first-generation quarks. Upper right: Mass dependence of the various decay modes of the scalar singlet $S$ induced by the $Q_{SqD}^{13}$ operator. Bottom: Mass dependence of the decay width of the scalar singlet $S$ for operators coupling the top quark to second-generation quarks. The results are shown setting $\Lambda_\text{NP} = 1\tev$ and all Wilson coefficients equal to one.}
  \label{fig:lightBSM_decays_S}
\end{figure}

The numerical results for the decay width of the scalar singlet $S$ (and the corresponding decay length) induced by operators coupling the top quark to first-generation quarks are shown in the upper left panel of \cref{fig:lightBSM_decays_S}. The decay width of $S$ strongly depends on the nature of the operator inducing the decay(s) as well as on the mass of $S$. 

Starting with the $Q_{SqD}^{13}$ operator --- using the superscript to denote the flavour of the involved quark fields ---, the $S$ is decaying promptly for $m_S >  m_b$ and long-lived for $m_S < m_b$. The contribution of the various decay channels is broken down in the upper right panel of \cref{fig:lightBSM_decays_S}. The tree-level decay to $b d$ dominates in the region of $m_S > m_B$. For $m_S < m_B$, the decay width is orders of magnitude smaller and originates from loop-induced decays where the top quark is traded for a first- or second-generation quark via the $W$ boson in the loop (receiving a suppression by the off-diagonal entries of the CKM matrix). The off-shell four-body decay into three light and one $b$ jet is only relevant for high $m_S$, for which at least the intermediary $W$ boson can be on-shell.

For the $Q_{SqD}^{31}$ operator, the tree-level decays are proportional to the down quark mass, which strongly suppresses it.\footnote{The derivative acting on $q_1$ can be rewritten in terms of the mass of the quark on which it is acting using the Dirac equation.} For the same reason, also the loop-induced and the four-body tree-level decays are strongly suppressed.

For the Yukawa-type operators ($Q_{Su\Phi}^{13}$ and $Q_{Su\Phi}^{31}$), there are no tree-level two-body decays and the tree-level four-body decays give a sizeable rate only for large $m_S$. For the $Q_{Su\Phi}^{31}$ operator, also the loop-induced decay channels are suppressed due to the chirality structure of the operator resulting in $S$ being long-lived for large parts of the considered parameter space. For the $Q_{Su\Phi}^{13}$ operator, on the other hand, the loop-induced decay channels are sizeable resulting in $S$ being long-lived only for $m_S\lesssim m_b$.

In the lower panel of \cref{fig:lightBSM_decays_S}, we look at the changes in the $S$ decay rate if it couples the top quark to a second-generation quark. In comparison to first-generation operators in the upper panels of \cref{fig:lightBSM_decays_S}, the main changes are visible for the $Q_{SqD}^{32}$ and $Q_{Su\Phi}^{32}$ operators. For the $Q_{SqD}^{32}$ operator, both the tree-level and one-loop two-body decays are significantly enhanced due to the strange-quark mass being substantially larger than the down-quark mass. For the same reason, the loop-level two-body decay induced by the $Q_{Su\Phi}^{32}$ operator is significantly enhanced. In general, the possibility of $S$ being long-lived is more constrained if it couples the top quark to second-generation quarks instead of first-generation quarks. This conclusion also holds for the fermionic and vector singlets $N$ and $S$. Therefore, we in the following only discuss the case in which they couple the top quark to first-generation quarks.

\begin{figure}
  \centering
  \includegraphics[width=.48\textwidth]{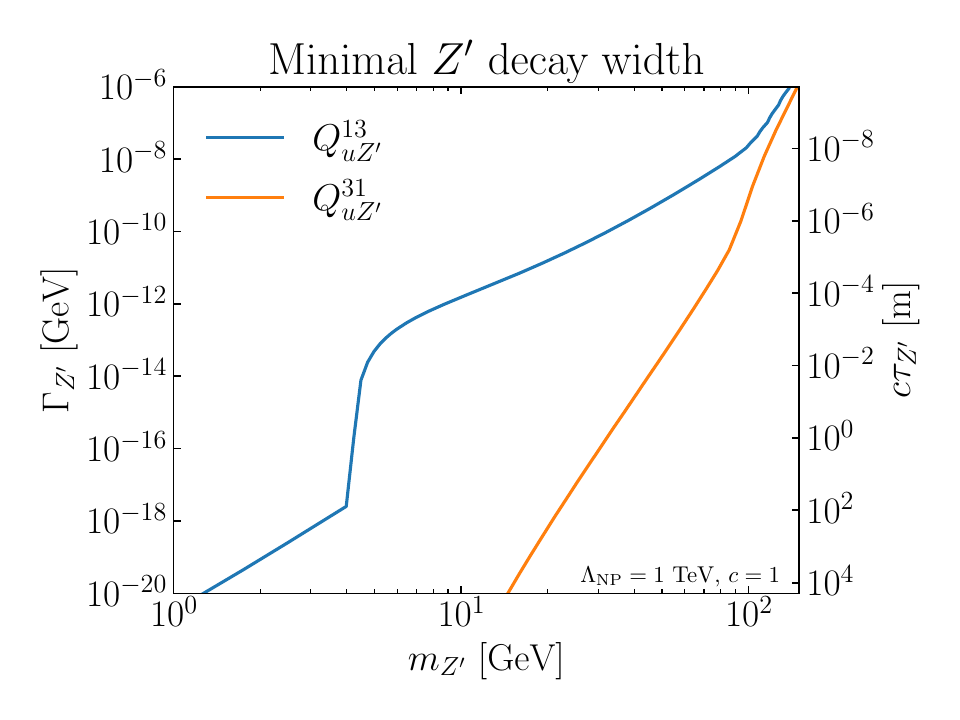}
  \includegraphics[width=.48\textwidth]{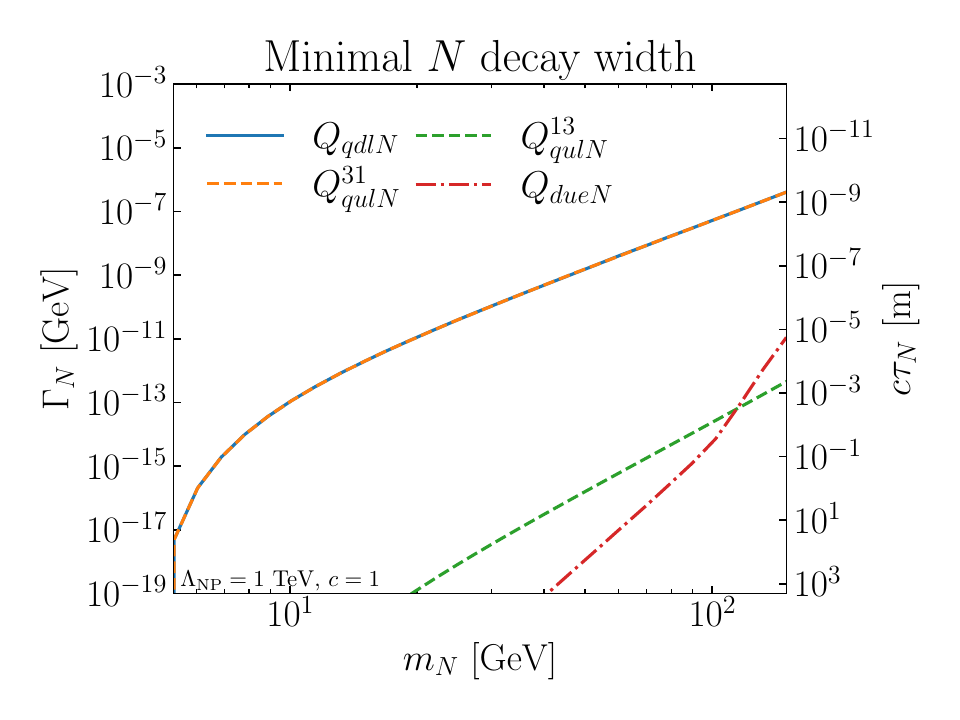}
  \caption{Mass dependence of the decay widths of the gauge boson $Z^\prime$ (left panel), and the sterile neutrino $N$ (right panel) into SM particles. The results are shown setting $\Lambda_\text{NP} = 1\tev$ and all Wilson coefficients equal to one.}
  \label{fig:lightBSM_decays_Zp_N}
\end{figure}

In the case of the singlet vector $Z'$ (see left panel of \cref{fig:lightBSM_decays_Zp_N}), the nature of its decay depends strongly on the flavour structure of the dipole operator. In the case of $Q_{uZ^\prime}^{13}$, the two-body loop-induced decay dominates for lower $Z^\prime$ masses whereas for higher masses the tree-level four-body decay gives a sizeable contribution. These decays are prompt whenever the $Z^\prime$ mass is sufficiently above the bottom-quark mass. However, for small enough $Z'$ masses such that the only available channels are to light quarks, these decays have additional CKM suppression which can make the $Z'$ long-lived. In fact in the benchmark point of \cref{fig:lightBSM_decays_Zp_N} it transitions quickly to being so long-lived as to be missing energy, but in detail, this shape will depend on the size of the coupling. In the case of $Q_{uZ^\prime}^{31}$, the loop-induced decay is strongly suppressed due to the chirality structure of the operator making $Z^\prime$ long-lived if $m_{Z^\prime} \lesssim 80\gev$. For higher masses, the tree-level four-body decay gives a substantial contribution.

For the right-handed neutrino $N$, its decays are further suppressed by the three-body phase space required with only the minimal four-fermion operators we have added. This makes a fermion singlet a natural candidate for being long-lived. The size of the $N$ decay width, however, depends sensitively on the structure of the operator inducing the top-quark decay (see right panel of \cref{fig:lightBSM_decays_Zp_N}). The \OqdlN and $\OqulN^{31}$ operators induce tree-level three-body decays resulting in prompt decays once $m_N\gtrsim 10\gev$. In the case of the $\OqulN^{13}$ and \OdueN operators no tree-level three-body decays exist. Moreover, also the loop-induced three-body decays are strongly suppressed due to the chirality structure of the operators. This makes the five-body tree-level decays the leading decay channels resulting in $N$ being long-lived throughout almost the complete considered parameter region.

\begin{figure}
  \centering
  \includegraphics[width=.8\textwidth]{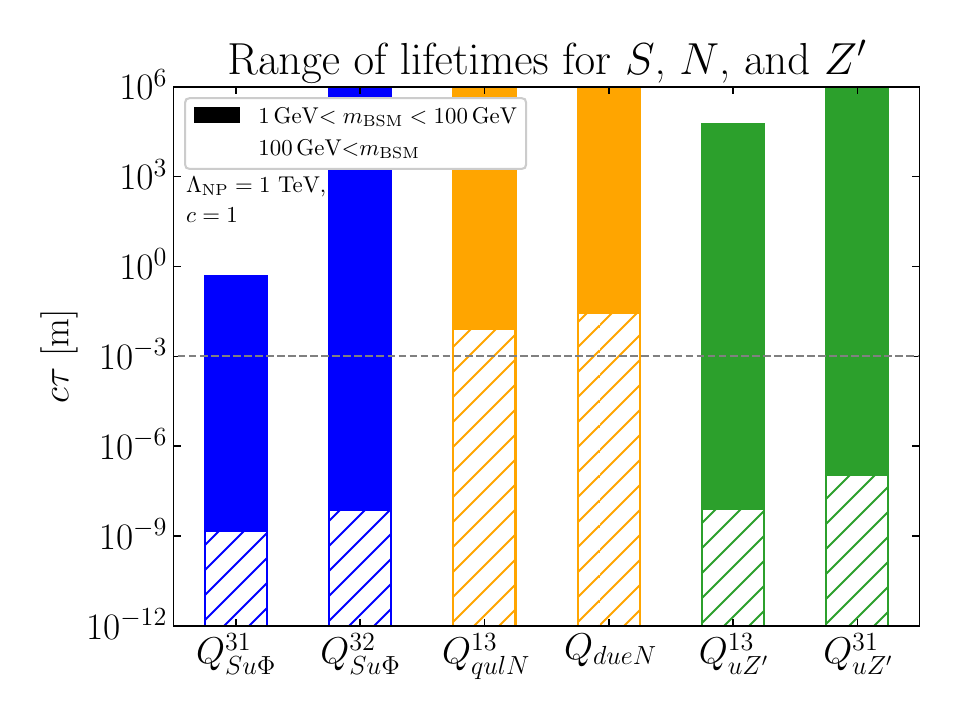}
  \caption{Lifetimes of the singlets $S$ (blue), $N$ (orange), and $Z^\prime$ (green) interacting through the indicated operators over a range of singlet masses. The longest lifetimes are achieved for $m_\text{BSM}=1 \gev$ at the upper end (in some cases they exceed the upper limit of the plot); the hatched regions denote masses larger than $100\gev$. Lifetimes well below the dashed line at $1 \text{mm}$ produce prompt decays rather than displaced ones.}
  \label{fig:BSM_decay_summary}
\end{figure}

This discussion clearly shows that the minimal decay width of the light singlets depends strongly on their mass. In summary, the following collider signatures arise in the
absence of other non-top operators:
\begin{itemize}
    \item \textbf{$m_\text{BSM} \lesssim m_b$} --- If the singlet mass is below the bottom-quark mass, the singlet is likely long-lived. For $S$ and $Z^\prime$, the resulting collider signature is a pair of jets originating from a displaced vertex. Similarly, for $N$, three jets originate from a displaced vertex.
    \item \textbf{$m_\text{BSM} > m_b$} --- If the singlet mass is above the bottom-quark mass, the singlet can be long-lived or decay promptly depending mainly on the chirality structure of the involved operator. For $S$ and $Z^\prime$, the collider signature is either a di-jet or a di-jet plus $W$ boson final state originating potentially from a displaced vertex. For $N$, one additional jet is present in the final state.
\end{itemize}
To provide an overview, we show in \cref{fig:BSM_decay_summary} the range of lifetimes for the singlets $S$, $N$, and $Z^\prime$ for the various operators (leaving out those operators which induce tree-level $B$ meson decays, see \cref{sec:complementary_constraints}). As discussed, the longest lifetimes are expected for the singlet fermion $N$.

While we have been focusing on the minimal cases, the signal for the LLP can in principle be much richer, as the singlet could have other decay modes in addition to those dictated by the operators involved in the rare top-quark decays. This is particularly interesting in cases where the minimal decay width of the new species is very small, as adding new decay channels with comparable partial widths does not dramatically change the decay lifetime. For example, light right-handed neutrinos $N$ minimally coupled to the top quark may be so long-lived that they appear as missing energy at the LHC. But of course, the $N$ may interact via the relevant `neutrino portal' $y_N \Phi L_L N$ or additional higher-dimensional operators $(Q_{L1} \bar d_{R1})(L_{L} \bar N)/\Lambda^2$ which reduce its lifetime and provide distinctive long-lived $N \rightarrow e^+ + \pi^-$ decays in the tracker. For the scalar, a small coupling $S F^{\mu\nu} F_{\mu\nu}$ (or the $\mathcal{CP}$-odd version) can lead to displaced photon signatures for which search strategies have been developed \cite{CMS:2019zxa,ATLAS:2022vhr}.


\subsection{Complementary constraints}
\label{sec:complementary_constraints}

Our goal in this Section is to outline the relevant constraints arising from low-energy probes of flavor physics. Our perspective is that the FCNC operators we study for rare top decays will be one contribution to these flavor signatures, but there may well be others, and it is possible in a full model there may be cancellations in contributions to any particular observable. While we will discuss the signatures of various operators in flavor observables, and indeed it is possible for them to be quite constraining, none of this dampens the usefulness of direct searches for the complementary rare top decay signatures. Since we are here concerned with these effective operators in particular rather than in constraining complete models of UV physics which include them, we will only perform a qualitative analysis since a precise evaluation is beyond the scope of the current work and would likely also depend on the concrete UV scenario.

First, let us briefly address the preference of this new physics for top quarks: From the point of view of the gauge quantum numbers of the Standard Model particles, there is no reason for some new singlet to couple to some particular flavors, or for that matter to prefer quarks to leptons. One might then worry from this perspective whether our consideration of such `top-philic' operators comprises some severe theoretical fine-tuning. At the same time, the top quark is special in the flavor structure of the Standard Model. It is also expected to play a crucial role in the dynamics of electroweak symmetry breaking. If the related new physics to address these questions is also responsible for rare decays into singlets, a preference for the top quark can then be understood. While we do not embed our effective operators in full ultraviolet models, we can have in the back of our minds simple scenarios where such a structure does come about. For example, at higher energies these singlets may be charged under a flavored gauge symmetry that has been spontaneously broken at energies above $\Lambda_\text{NP}$, such as gauged horizontal baryon number $U(1)_{B_1 - B_3}$ or third-generation baryon minus lepton number $U(1)_{(B-L)_3}$. For further such considerations we refer to, for example, \ccite{Alonso:2017uky,Babu:2017olk,Alonso:2017bff,Bordone:2017bld,Greljo:2018tuh,Fox:2018ldq,Elahi:2019drj}, among many such works.

\begin{figure}
  \centering
    \includegraphics[width=.42\textwidth]{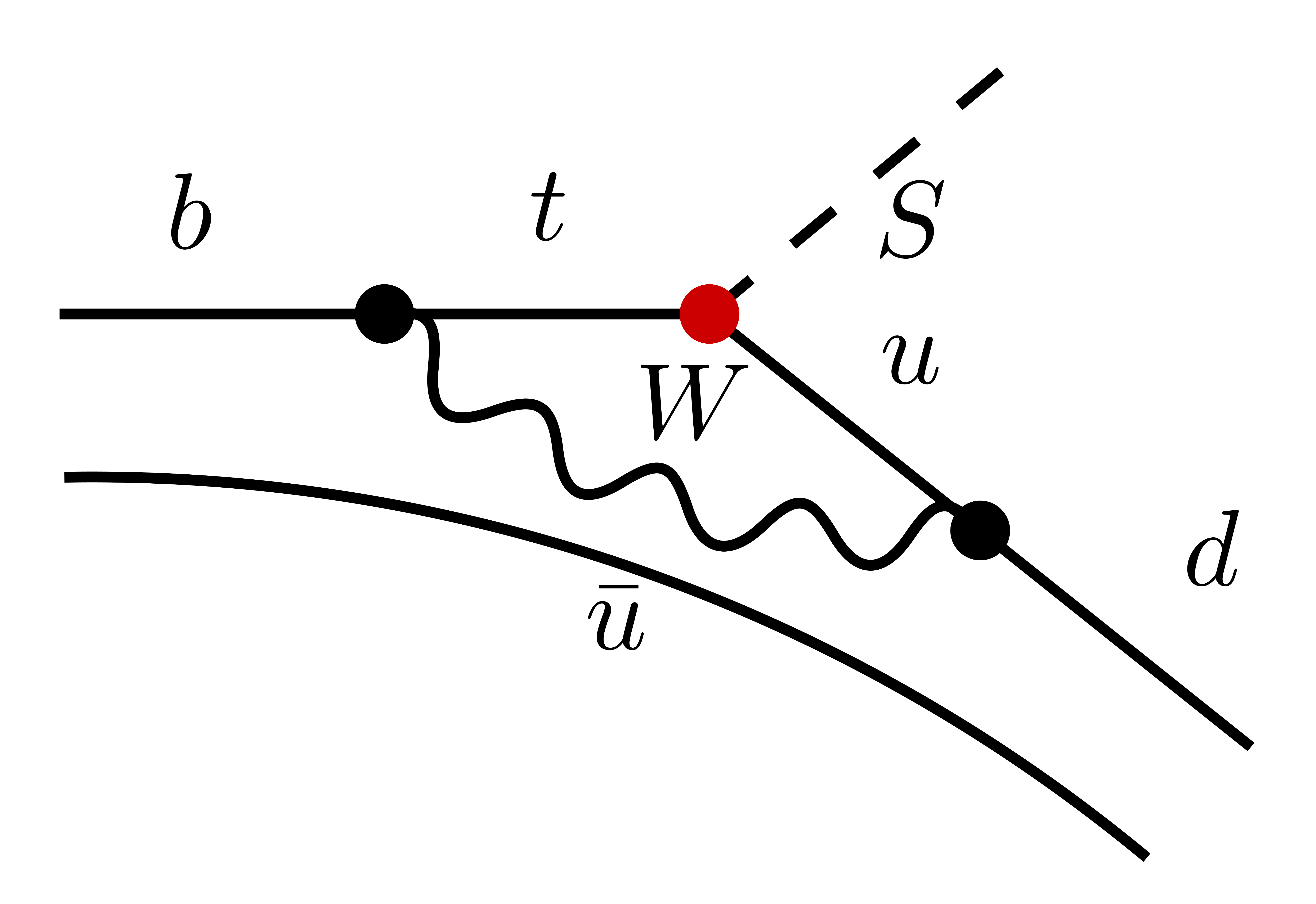}
  \includegraphics[width=.48\textwidth]{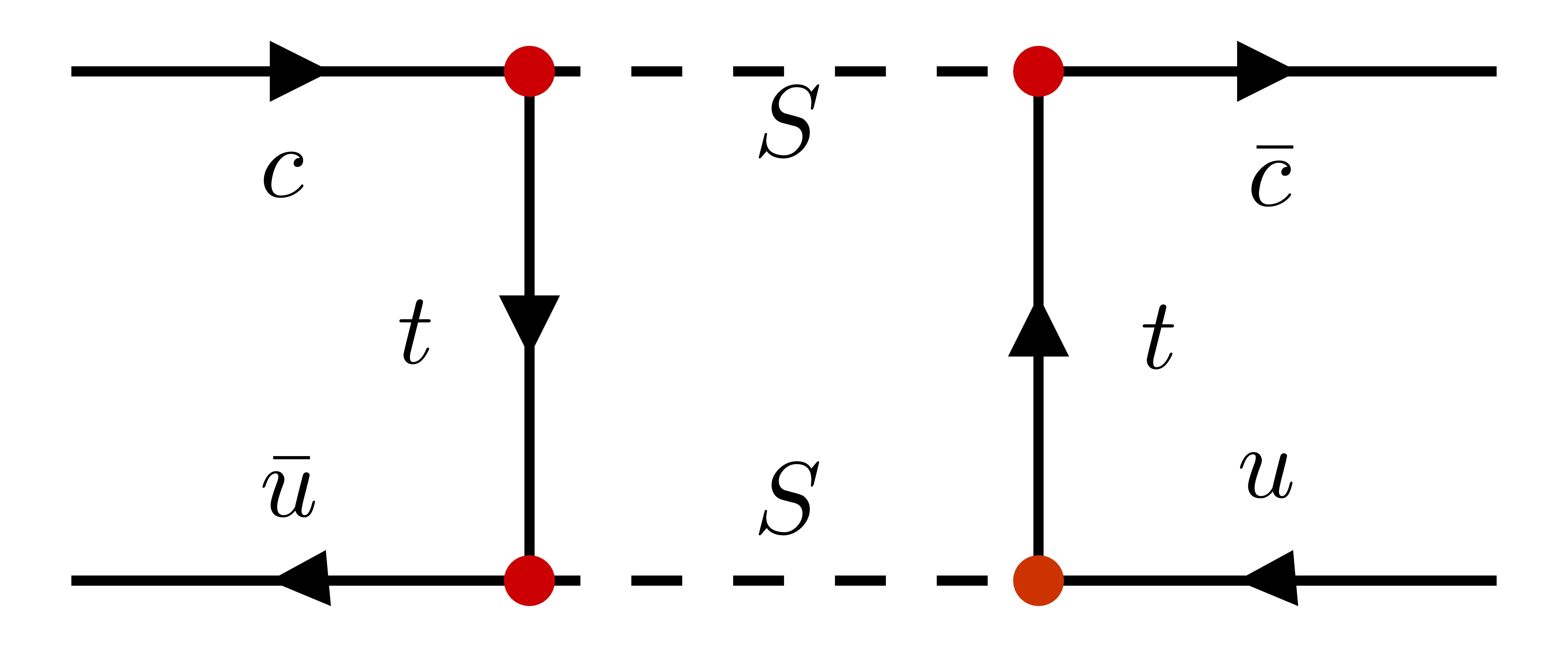}
    \includegraphics[width=.48\textwidth]{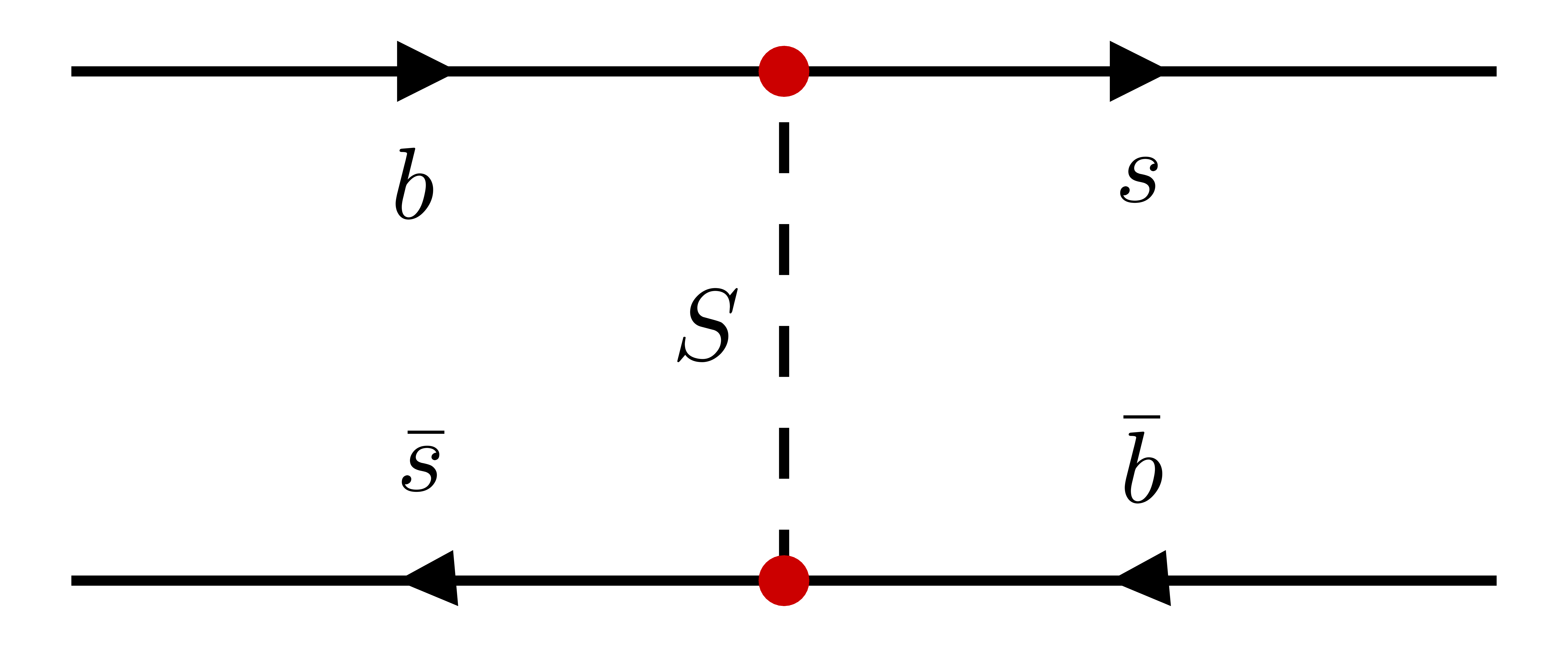}
  \includegraphics[width=.42\textwidth]{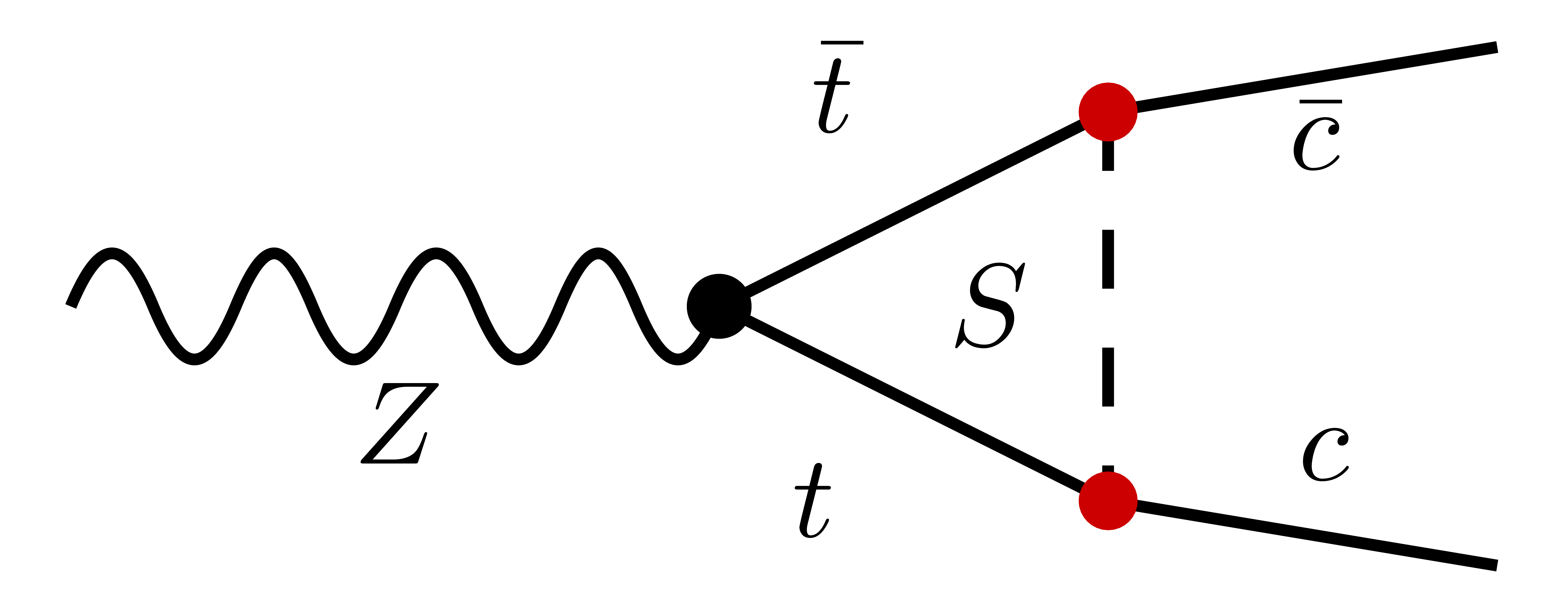}
  \caption{Sample Feynman diagrams for (upper left) loop-induced decays of $B$ mesons to missing energy; (upper right) loop-induced mixing of $D$ mesons; (lower left) tree-level mixing of $B_s$ mesons; (lower right) loop-level decays of $Z$ bosons.}
  \label{fig:flavorDiags}
\end{figure}

Given that our new physics is manifestly flavour-violating, there can be strong constraints from the effects of flavor-changing neutral currents. This is most pertinent for the scalar $S$ which has flavor-changing interactions at dimension-five, so we discuss this possibility first. The possible effects naturally divide into four cases depending upon the mass and couplings: 

\begin{itemize}
    \item \textbf{Sub-mesonic scalar} --- When the new singlet species is light enough (e.g., below the $B$-meson mass scale $m_B$), it may be produced in meson decays \cite{Kamenik:2011vy}. 
    Contributions to meson decays e.g. $B^+ \rightarrow \pi^+ + S$ can appear as diagrammed in the upper left of \cref{fig:flavorDiags}. 
    With $\mathcal{O}(1)$ coupling, the decay width can be far larger than in the SM where it proceeds through an off-shell $W$ into a pair of fermions and is suppressed by phase space and an off-diagonal CKM.
    In fact, such exotic decays are probed by searches for decays such as $B^{\pm} \rightarrow \pi^\pm \nu \bar \nu$, which are constrained now down to a branching ratio below $\sim 10^{-5}$, and similarly for decays to kaons \cite{Belle:2017oht}. 
    Resultingly, extremely light singlet bosons with flavor-changing dimension-5 interactions must have quite suppressed couplings, though in detail the constraint will depend on which operator is turned on along the lines of the $S$ lifetime discussion. 
    If the bottom coupling is present already at tree-level, as with $(c/\Lambda_{NP}) S(\bar Q_{Li} \slashed{D} Q_{Lj})$, this gives a strong constraint which is roughly $c^2 (1 \tev/\Lambda_{NP})^2 \lesssim 10^{-11}$. 
    If instead there is a coupling to light right-handed quarks, as for $S(\bar Q_{L3} u_{R1} \tilde{\Phi})$, a small Yukawa coupling is needed to flip the chirality and allow a $W$ loop. Alternatively, the decay may happen at the tree level through an off-shell $W$, for example, $B^0 \rightarrow S + \pi^- + e^+ + \nu_e$, which would appear as extra contributions to the SM $\text{Br}(B^0 \rightarrow \pi^- + e^+ + \nu_e)$ which has been measured at $\sim 10^{-4}$ \cite{Belle:2013hlo}.\footnote{We note that the application of these constraints does assume the $S$ is invisible, and so one could imagine adding further operators allowing even a light $S$ to decay to mesons or charged leptons and hide amongst the uncertainties in those decay modes.} Relative to the SM decay, which is suppressed by an off-diagonal CKM element, the BSM contribution is suppressed by an additional off-shell top quark, the inserted top-decay operator, as well as the phase space volume (due to the additional particle in the final state). Consequently, any reasonable BSM contribution should be significantly smaller than the experimental uncertainty of $\sim 10^{-5}$.

    \item \textbf{Flavor-anarchic scalar} --- If the new light species couples to all three generations (that is, if both $t$--$u$--$S$ and $t$--$c$--$S$ couplings exist), then a loop of top quarks and the $S$ can induce oscillations among flavored neutral mesons, such as in the upper right diagram of \cref{fig:flavorDiags}. Such effects in the SM appear at one loop and are suppressed by a small CKM angle $V_\text{small}$ and the electroweak scale. These mixing phenomena have been measured (e.g. \ccite{LHCb:2021ykz}) and agree with the SM predictions, so at the very least our new contributions must be subdominant. The effect appears here with the same diagram topology, so the scale of our operator needs only be mildly higher to overcome to the CKM suppression, very roughly $\Lambda_{NP}^4 \gg v_{EW}^4/V_\text{small}^2$. Since the top quark branching ratio scales as $\Lambda_\text{NP}^2$, this is not too significant a constraint on exotic decays into $S$.  

    \item \textbf{Non-chiral scalar} --- If the scalar couples both to up-type and down-type quarks, as with \OSqD, then tree-level exchanges can lead to $B^0$ or $B_s^0$ oscillations, as in the lower left diagram of \cref{fig:flavorDiags}. The SM predictions for these phenomena are in good agreement with the data (see e.g.\ \ccite{LHCb:2016gsk}). Here the new effect appears at tree-level, so is significantly enhanced, and the requirement of being smaller than the SM rates is roughly $\Lambda_\text{NP}^2 \gg v_\text{EW}^4/(m_B^2 V_\text{small}^2)$, which is far more severe a constraint on the rate of exotic top quark decays.
    These effects do not appear if we consider instead the operator \OSuP without the analogous down-quark Yukawa-like operator, as can be enforced by $S$ being charged under a Peccei-Quinn symmetry.

    \item \textbf{Flavorful, up-philic scalar} --- If the new physics couples the third generation to only one of the light generations, and only to the up-type quarks, the prior effects are absent. Then the most relevant effects arise from closing the top-quark loop with a $Z$ boson coupling, and the resulting constraint is from affecting electroweak precision. For example, at one loop there is a new contribution to the width of the $Z$ boson shown in the lower right diagram of \cref{fig:flavorDiags}. While the $Z$ boson width has been measured quite precisely (see e.g.\ \ccite{ALEPH:2005ab}), this contribution is easily reduced to below the level of the many SM 1-loop diagrams. 
\end{itemize}
Let us emphasize this safest case, having discussed the different sorts of flavor effects which can constrain $S$ depending on its mass and the structure of its interactions. For a scalar with mass $m_S > m_B$ coupled to the top quark through $S(\bar Q_{L_i} u_{Rj} \tilde \Phi)$, the complementary constraints are ineffective and these scalars are best searched for with rare top decays. If $S$ couples only to left-handed top quarks and either a right-handed up or charm quark ($i=3$ and $j=1$ or $j=2$), $S$ is also long-lived up to masses $\sim 50 \gev$, meaning these searches can be especially sensitive. In this case, also for $m_S < m_B$ flavour constraints are very weak. These coupling structures can appear naturally if the scalar is charged under a flavorful Peccei-Quinn symmetry.

In the case where $S$ is endowed with a $\mathbb{Z}_2$ symmetry such that it couples only in pairs, the same diagrams appear but with an extra scalar leg. This leads to additional $\Lambda_\text{NP}^{-2}$ suppression from having a higher-dimensional operator, as well as additional loop suppression from closing the extra line. The constraints in this case are thus much weaker.

Similarly, for a singlet vector $Z'$, the same diagrams are present as in the case of a single $S$, so approximately similar considerations apply. Since the leading operator is at dimension six, there is additional suppression from each BSM vertex of $m_t/\Lambda_\text{NP}$ or $m_{Z'}/\Lambda_\text{NP}$, which softens the constraints relative to the scalar case. Again the least stringently constrained option is for a massive $Z'$ which couples solely to up-type quarks and only to one of the light generations.

\begin{figure}
  \centering
    \includegraphics[width=.42\textwidth]{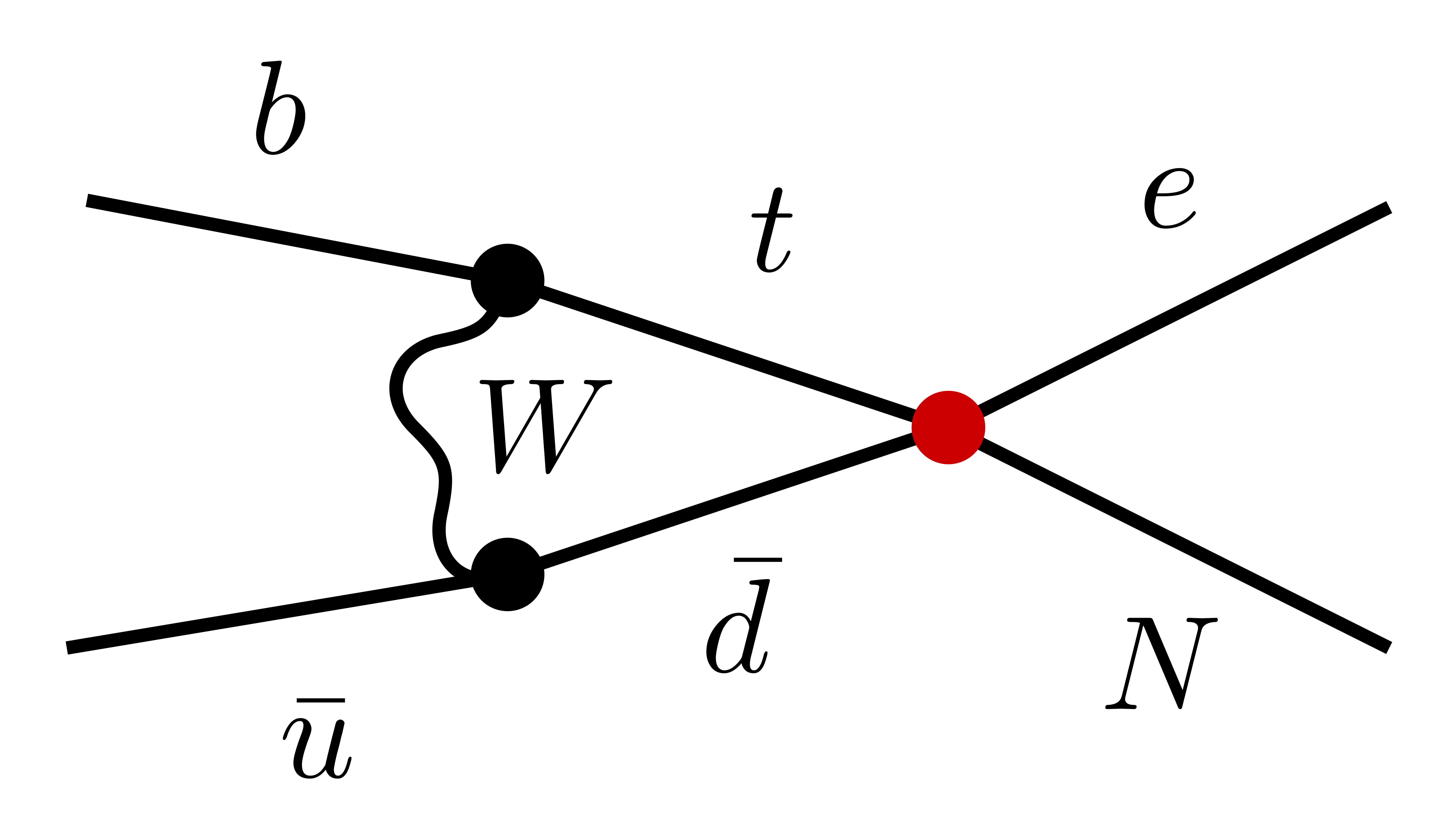}
  \includegraphics[width=.48\textwidth]{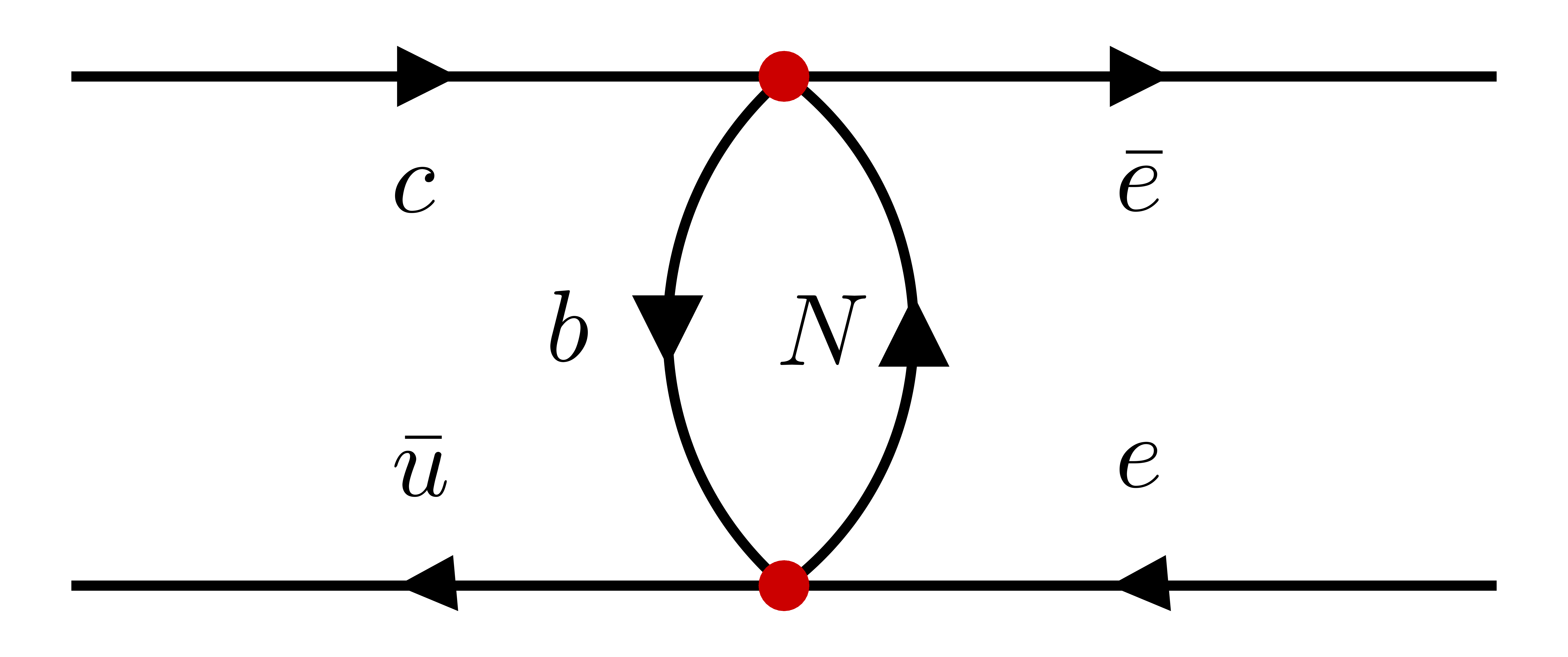}
    \includegraphics[width=.48\textwidth]{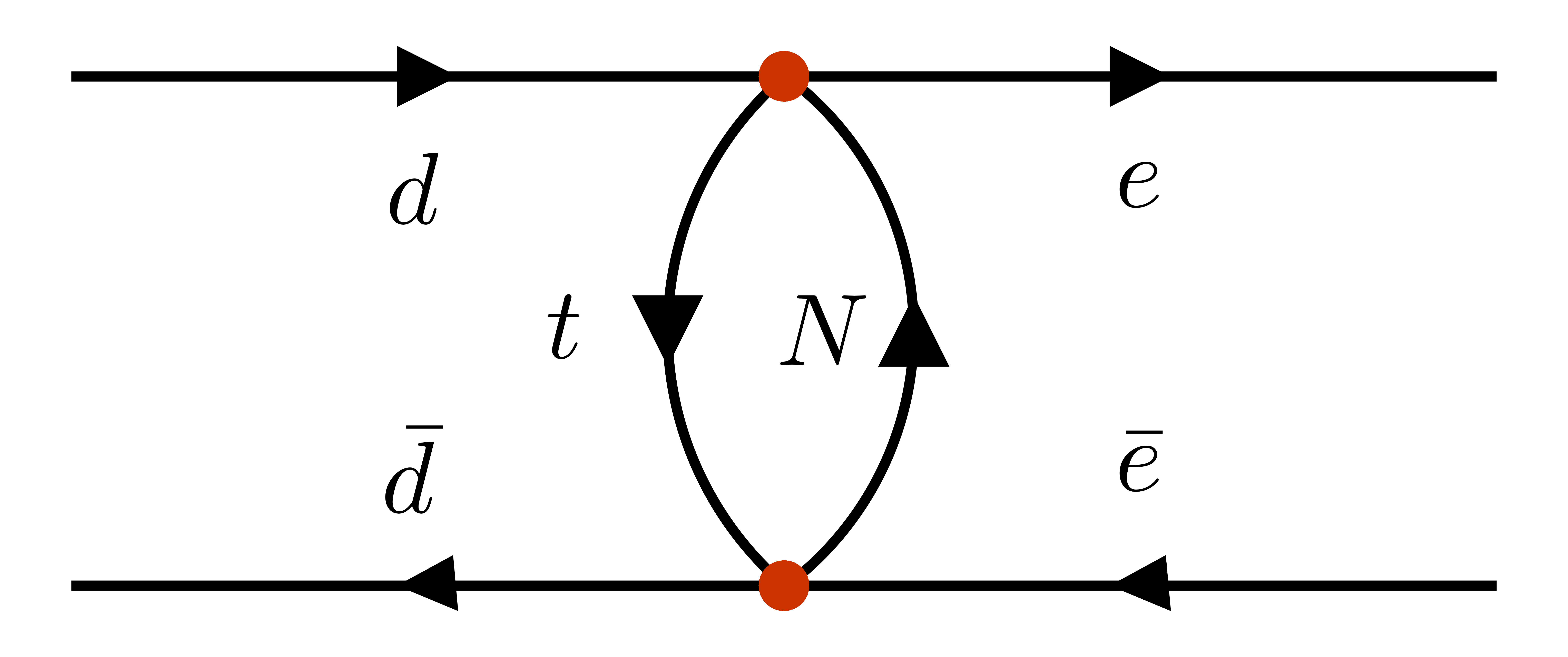}
  \includegraphics[width=.42\textwidth]{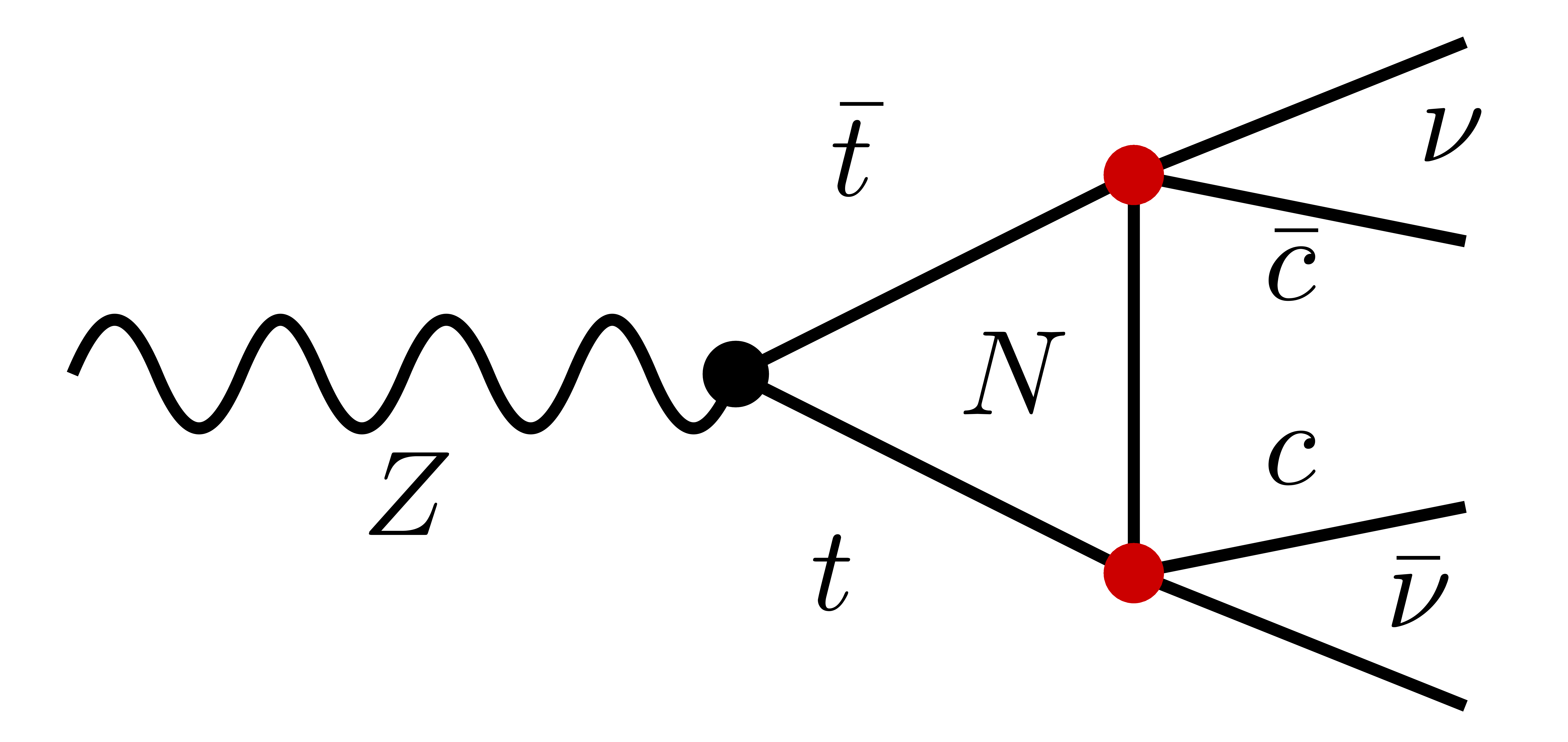}
  \caption{Sample Feynman diagrams for complementary effects of top decay operators with a singlet fermion $N$, including (upper left) leptonic decays of $B$ mesons; (upper right) decays of $D$ mesons; (lower left) decays of pions; (lower right) semi-visible decays of $Z$ bosons.}
  \label{fig:flavorDiagsN}
\end{figure}

The singlet fermion $N$ can lead to a similar variety of flavor-changing phenomena, though here the primary effects are in leptonic decays rather than in oscillations. The lowest operators are again at dimension six, making them relatively safe compared to the impact of a light scalar singlet. 

For light-enough $m_N \ll m_b$, one may have $W$-loop decays of $B$-mesons to $e + N$ or fully invisible $\nu + N$, as in the upper left diagram of \cref{fig:flavorDiagsN}. Such modes have been searched for and are constrained at the level of $\mathcal{O}(10^{-6})$ \cite{Belle:2006tbq} in the branching ratio. Since the SM width is suppressed by phase space and CKM, this implies a tight constraint on such light fermion couplings. Similarly to the scalar considerations, if the operator is the fully right-handed \OdueN, then instead the dominant induced decay mode may be at tree-level with an off-shell $W$, as $B^0 \rightarrow \pi^- + e^+ + N$. This decay mode will, however, be again significantly suppressed relative to the corresponding SM decay mode due to the off-shell top propagator, the inserted top-decay operator, and the additional particle in the final state.

Moving to heavier species, for fully flavored couplings, a loop of $t$ and $N$ leads to new leptonic $D$-meson decays, as in the upper right diagram of \cref{fig:flavorDiagsN}). These are tightly constrained to a branching ratio of $\lesssim 10^{-7}$ \cite{Belle:2010ouj}, though the new physics effect here scaling as $\Lambda_\text{NP}^{-8}$ makes this constraint not so severe. Even with only one flavor of coupling there are still decays of quarkonia to $\ell^+\ell^-$ (see lower left diagram of \cref{fig:flavorDiagsN}). Especially dangerous would be couplings to the first generation which give extra leptonic decay of pions, but even $J/\psi \rightarrow e^+e^-$ is measured quite precisely \cite{KEDR:2018lhb}. Again the faster scaling of this process with $\Lambda_\text{NP}$ compared to the top decays makes this less severe, but it is still an important consideration. The additional contribution to the $Z$ boson width is in this case further suppressed either by phase space (as in the lower right diagram of \cref{fig:flavorDiagsN}) or another loop.

In summary, low-energy flavour and electroweak constraints can be sensitive to a subset of top-decay operators. The possible BSM final states shown in \cref{fig:summary} can, however, all\footnote{With the exception of the $t\to SZj,S\gamma j, SWj$ channels.} be realized without being constrained by flavour or electroweak precision measurements.


\section{Conclusion} \label{sec:conclusion}

The Large Hadron Collider, having achieved in its infancy huge success as a discovery machine, is maturing into a precision machine for its next decades of cutting-edge exploration into the structure of matter. Rare decays of top quarks offer an exciting window into potential deviations from Standard Model physics, either with flavor-changing effects or in the observation of new light singlets. 

To set up the framework of our study, we parameterized the interaction relevant to the rare decay of the top quark by a set of EFT operators, including all pertinent ones up to dimension six.  We then described the FCNC decay final states, including the ones with only SM particles, and those including an additional singlet. In both of these cases, there are channels with a large number of signal events at the HL-LHC, leading to great discovery potential. While we focused only on rare top decays, single top quark production offers additional sensitivity.

We then concentrated on the case of light singlet scalars, fermions, or vector bosons. To get an overview of the possible experimental signatures, we distinguished three different cases: promptly decaying singlets, long-lived singlets, and stable singlets. We discussed under which conditions each of these possibilities can be realized. Depending on the concrete operator inducing the exotic top-quark decay, the light singlets can be long-lived resulting in interesting experimental signatures. We also qualitatively discussed complementary constraints from flavor physics and electroweak precision measurements. While these can provide strong constraints on particular operators, they do not exclude any of the rare top decay signatures. More generally, the flavour and electroweak constraints should be considered only as a guide, since the full UV model may well contain additional BSM particles contributing to these observables.

In this initial work, our intent has been to point to these opportunities in general, and there are many avenues for future work. As highlighted above, the large number of rare FCNC top decays possible with dimension-six operators raises the prospect of even higher-dimensional operators still being usefully probed. It would be useful to understand more precisely all the possible final state signatures, and to ensure that experimental searches exist for all the useful cases. Finally, it would also be interesting to make further contact with particular ultraviolet models where one could understand in more detail the effects of complementary constraints from flavor physics and other possible correlated signatures.


\section*{Acknowledgments}

HB acknowledges support from the Alexander von Humboldt foundation. SK is supported by an Oehme Postdoctoral Fellowship from the Enrico Fermi Institute. LTW is supported by the DOE grant DE-SC0013642. The Feynman diagrams shown in this article have been generated using \texttt{FeynGame}~\cite{Harlander:2020cyh}.


\clearpage
\printbibliography

@article{Carmona:2022jid,
    author = "Carmona, Adrian and Elahi, Fatemeh and Scherb, Christiane and Schwaller, Pedro",
    title = "{The ALPs from the top: searching for long lived axion-like particles from exotic top decays}",
    eprint = "2202.09371",
    archivePrefix = "arXiv",
    primaryClass = "hep-ph",
    reportNumber = "MITP/22-017",
    doi = "10.1007/JHEP07(2022)122",
    journal = "JHEP",
    volume = "07",
    pages = "122",
    year = "2022"
}

@article{Grzadkowski:2010es,
    author = "Grzadkowski, B. and Iskrzynski, M. and Misiak, M. and Rosiek, J.",
    title = "{Dimension-Six Terms in the Standard Model Lagrangian}",
    eprint = "1008.4884",
    archivePrefix = "arXiv",
    primaryClass = "hep-ph",
    reportNumber = "IFT-9-2010, TTP10-35",
    doi = "10.1007/JHEP10(2010)085",
    journal = "JHEP",
    volume = "10",
    pages = "085",
    year = "2010"
}

@article{Kumar:2013hfa,
    author = "Kumar, Abhishek and Tulin, Sean",
    title = "{Top-flavored dark matter and the forward-backward asymmetry}",
    eprint = "1303.0332",
    archivePrefix = "arXiv",
    primaryClass = "hep-ph",
    reportNumber = "MCTP-13-04",
    doi = "10.1103/PhysRevD.87.095006",
    journal = "Phys. Rev. D",
    volume = "87",
    number = "9",
    pages = "095006",
    year = "2013"
}

@article{Renner:2018fhh,
    author = "Renner, Sophie and Schwaller, Pedro",
    title = "{A flavoured dark sector}",
    eprint = "1803.08080",
    archivePrefix = "arXiv",
    primaryClass = "hep-ph",
    doi = "10.1007/JHEP08(2018)052",
    journal = "JHEP",
    volume = "08",
    pages = "052",
    year = "2018"
}

@article{Altmannshofer:2019ogm,
    author = "Altmannshofer, Wolfgang and Maddock, Brian and Tuckler, Douglas",
    title = "{Rare Top Decays as Probes of Flavorful Higgs Bosons}",
    eprint = "1904.10956",
    archivePrefix = "arXiv",
    primaryClass = "hep-ph",
    doi = "10.1103/PhysRevD.100.015003",
    journal = "Phys. Rev. D",
    volume = "100",
    number = "1",
    pages = "015003",
    year = "2019"
}

@article{Dimopoulos:1979es,
    author = "Dimopoulos, Savas and Susskind, Leonard",
    editor = "Zichichi, A.",
    title = "{Mass Without Scalars}",
    reportNumber = "CU-TP-147, ITP-626-STANFORD",
    doi = "10.1016/0550-3213(79)90364-X",
    journal = "Nucl. Phys. B",
    volume = "155",
    pages = "237--252",
    year = "1979"
}

@article{Broggio:2019ewu,
    author = "Broggio, Alessandro and Ferroglia, Andrea and Frederix, Rikkert and Pagani, Davide and Pecjak, Benjamin D. and Tsinikos, Ioannis",
    title = "{Top-quark pair hadroproduction in association with a heavy boson at NLO+NNLL including EW corrections}",
    eprint = "1907.04343",
    archivePrefix = "arXiv",
    primaryClass = "hep-ph",
    reportNumber = "TUM-HEP-1208/19, LU-TP 19-30, IPPP/19/57",
    doi = "10.1007/JHEP08(2019)039",
    journal = "JHEP",
    volume = "08",
    pages = "039",
    year = "2019"
}

@article{Gripaios:2009pe,
    author = "Gripaios, Ben and Pomarol, Alex and Riva, Francesco and Serra, Javi",
    title = "{Beyond the Minimal Composite Higgs Model}",
    eprint = "0902.1483",
    archivePrefix = "arXiv",
    primaryClass = "hep-ph",
    doi = "10.1088/1126-6708/2009/04/070",
    journal = "JHEP",
    volume = "04",
    pages = "070",
    year = "2009"
}

@book{Panico:2015jxa,
    author = "Panico, Giuliano and Wulzer, Andrea",
    title = "{The Composite Nambu-Goldstone Higgs}",
    eprint = "1506.01961",
    archivePrefix = "arXiv",
    primaryClass = "hep-ph",
    reportNumber = "DFPD-2015TH9",
    doi = "10.1007/978-3-319-22617-0",
    publisher = "Springer",
    volume = "913",
    year = "2016"
}

@article{Altmannshofer:2016zrn,
    author = "Altmannshofer, Wolfgang and Eby, Joshua and Gori, Stefania and Lotito, Matteo and Martone, Mario and Tuckler, Douglas",
    title = "{Collider Signatures of Flavorful Higgs Bosons}",
    eprint = "1610.02398",
    archivePrefix = "arXiv",
    primaryClass = "hep-ph",
    reportNumber = "FERMILAB-PUB-16-499-PPD",
    doi = "10.1103/PhysRevD.94.115032",
    journal = "Phys. Rev. D",
    volume = "94",
    number = "11",
    pages = "115032",
    year = "2016"
}

@article{Das:1995df,
    author = "Das, Ashok K. and Kao, Chung",
    title = "{A Two Higgs doublet model for the top quark}",
    eprint = "hep-ph/9511329",
    archivePrefix = "arXiv",
    reportNumber = "UR-1446",
    doi = "10.1016/0370-2693(96)00031-7",
    journal = "Phys. Lett. B",
    volume = "372",
    pages = "106--112",
    year = "1996"
}

@article{Altmannshofer:2015esa,
    author = "Altmannshofer, Wolfgang and Gori, Stefania and Kagan, Alexander L. and Silvestrini, Luca and Zupan, Jure",
    title = "{Uncovering Mass Generation Through Higgs Flavor Violation}",
    eprint = "1507.07927",
    archivePrefix = "arXiv",
    primaryClass = "hep-ph",
    doi = "10.1103/PhysRevD.93.031301",
    journal = "Phys. Rev. D",
    volume = "93",
    number = "3",
    pages = "031301",
    year = "2016"
}

@article{Kilic:2015vka,
    author = "Kilic, Can and Klimek, Matthew D. and Yu, Jiang-Hao",
    title = "{Signatures of Top Flavored Dark Matter}",
    eprint = "1501.02202",
    archivePrefix = "arXiv",
    primaryClass = "hep-ph",
    reportNumber = "UTTG-28-14, TCC-027-14",
    doi = "10.1103/PhysRevD.91.054036",
    journal = "Phys. Rev. D",
    volume = "91",
    number = "5",
    pages = "054036",
    year = "2015"
}

@article{Batell:2013zwa,
    author = "Batell, Brian and Lin, Tongyan and Wang, Lian-Tao",
    title = "{Flavored Dark Matter and R-Parity Violation}",
    eprint = "1309.4462",
    archivePrefix = "arXiv",
    primaryClass = "hep-ph",
    doi = "10.1007/JHEP01(2014)075",
    journal = "JHEP",
    volume = "01",
    pages = "075",
    year = "2014"
}

@article{Belle:2017oht,
    author = "Grygier, J. and others",
    collaboration = "Belle",
    title = "{Search for $\boldsymbol{B\to h\nu\bar{\nu}}$ decays with semileptonic tagging at Belle}",
    eprint = "1702.03224",
    archivePrefix = "arXiv",
    primaryClass = "hep-ex",
    doi = "10.1103/PhysRevD.96.091101",
    journal = "Phys. Rev. D",
    volume = "96",
    number = "9",
    pages = "091101",
    year = "2017",
    note = "[Addendum: Phys.Rev.D 97, 099902 (2018)]"
}

@article{Alonso:2017bff,
    author = "Alonso, Rodrigo and Cox, Peter and Han, Chengcheng and Yanagida, Tsutomu T.",
    title = "{Anomaly-free local horizontal symmetry and anomaly-full rare B-decays}",
    eprint = "1704.08158",
    archivePrefix = "arXiv",
    primaryClass = "hep-ph",
    reportNumber = "IPMU17-0068, CERN-TH-2017-094",
    doi = "10.1103/PhysRevD.96.071701",
    journal = "Phys. Rev. D",
    volume = "96",
    number = "7",
    pages = "071701",
    year = "2017"
}

@article{Bordone:2017bld,
    author = "Bordone, Marzia and Cornella, Claudia and Fuentes-Martin, Javier and Isidori, Gino",
    title = "{A three-site gauge model for flavor hierarchies and flavor anomalies}",
    eprint = "1712.01368",
    archivePrefix = "arXiv",
    primaryClass = "hep-ph",
    reportNumber = "ZU-TH-36-17",
    doi = "10.1016/j.physletb.2018.02.011",
    journal = "Phys. Lett. B",
    volume = "779",
    pages = "317--323",
    year = "2018"
}

@article{Greljo:2018tuh,
    author = "Greljo, Admir and Stefanek, Ben A.",
    title = "{Third family quark\textendash{}lepton unification at the TeV scale}",
    eprint = "1802.04274",
    archivePrefix = "arXiv",
    primaryClass = "hep-ph",
    reportNumber = "MITP-18-012",
    doi = "10.1016/j.physletb.2018.05.033",
    journal = "Phys. Lett. B",
    volume = "782",
    pages = "131--138",
    year = "2018"
}

@article{Babu:2017olk,
    author = "Babu, K. S. and Friedland, A. and Machado, P. A. N. and Mocioiu, I.",
    title = "{Flavor Gauge Models Below the Fermi Scale}",
    eprint = "1705.01822",
    archivePrefix = "arXiv",
    primaryClass = "hep-ph",
    reportNumber = "OSU-HEP-17-02, SLAC-PUB-16542, IFT-UAM-CSIC-16-051, FTUAM-16-21, FERMILAB-PUB-17-005-T, NSF-KITP-17-060",
    doi = "10.1007/JHEP12(2017)096",
    journal = "JHEP",
    volume = "12",
    pages = "096",
    year = "2017"
}

@article{Alonso:2017uky,
    author = "Alonso, Rodrigo and Cox, Peter and Han, Chengcheng and Yanagida, Tsutomu T.",
    title = "{Flavoured $B-L$ local symmetry and anomalous rare $B$ decays}",
    eprint = "1705.03858",
    archivePrefix = "arXiv",
    primaryClass = "hep-ph",
    reportNumber = "IPMU17-0075, CERN-TH-2017-106",
    doi = "10.1016/j.physletb.2017.10.027",
    journal = "Phys. Lett. B",
    volume = "774",
    pages = "643--648",
    year = "2017"
}

@article{Elahi:2019drj,
    author = "Elahi, Fatemeh and Martin, Adam",
    title = "{LHC constraints on a $(B-L)_3$ gauge boson}",
    eprint = "1905.10106",
    archivePrefix = "arXiv",
    primaryClass = "hep-ph",
    doi = "10.1103/PhysRevD.100.035016",
    journal = "Phys. Rev. D",
    volume = "100",
    number = "3",
    pages = "035016",
    year = "2019"
}

@article{Fox:2018ldq,
    author = "Fox, Patrick J. and Low, Ian and Zhang, Yue",
    title = "{Top-philic $Z'$ forces at the LHC}",
    eprint = "1801.03505",
    archivePrefix = "arXiv",
    primaryClass = "hep-ph",
    reportNumber = "FERMILAB-PUB-18-005-T, NUHEP-TH-18-01, CERN-TH-2018-010",
    doi = "10.1007/JHEP03(2018)074",
    journal = "JHEP",
    volume = "03",
    pages = "074",
    year = "2018"
}

@article{Chala:2018agk,
    author = "Chala, Mikael and Santiago, Jose and Spannowsky, Michael",
    title = "{Constraining four-fermion operators using rare top decays}",
    eprint = "1809.09624",
    archivePrefix = "arXiv",
    primaryClass = "hep-ph",
    reportNumber = "IPPP/18/82",
    doi = "10.1007/JHEP04(2019)014",
    journal = "JHEP",
    volume = "04",
    pages = "014",
    year = "2019"
}

@article{ATLAS:2017vgz,
    author = "Aaboud, Morad and others",
    collaboration = "ATLAS",
    title = "{Direct top-quark decay width measurement in the $t\bar{t}$ lepton+jets channel at $\sqrt{s}$=8 TeV with the ATLAS experiment}",
    eprint = "1709.04207",
    archivePrefix = "arXiv",
    primaryClass = "hep-ex",
    reportNumber = "CERN-EP-2017-187",
    doi = "10.1140/epjc/s10052-018-5595-5",
    journal = "Eur. Phys. J. C",
    volume = "78",
    number = "2",
    pages = "129",
    year = "2018"
}

@article{Knapen:2022afb,
    author = "Knapen, Simon and Lowette, Steven",
    title = "{A guide to hunting long-lived particles at the LHC}",
    eprint = "2212.03883",
    archivePrefix = "arXiv",
    primaryClass = "hep-ph",
    month = "12",
    year = "2022"
}

@article{CMS:2019zxa,
    author = "Sirunyan, Albert M and others",
    collaboration = "CMS",
    title = "{Search for long-lived particles using delayed photons in proton-proton collisions at $\sqrt{s}=$ 13 TeV}",
    eprint = "1909.06166",
    archivePrefix = "arXiv",
    primaryClass = "hep-ex",
    reportNumber = "CMS-EXO-19-005, CERN-EP-2019-185",
    doi = "10.1103/PhysRevD.100.112003",
    journal = "Phys. Rev. D",
    volume = "100",
    number = "11",
    pages = "112003",
    year = "2019"
}

@article{Bradshaw:2023wco,
    author = "Bradshaw, Layne and Chang, Spencer",
    title = "{Primary Observables for Top Quark Collider Signals}",
    eprint = "2304.06063",
    archivePrefix = "arXiv",
    primaryClass = "hep-ph",
    month = "4",
    year = "2023"
}

@article{Drobnak:2008br,
    author = "Drobnak, Jure and Fajfer, Svjetlana and Kamenik, Jernej F.",
    title = "{Signatures of NP models in top FCNC decay t ---\ensuremath{>} c(u) l+ l-}",
    eprint = "0812.0294",
    archivePrefix = "arXiv",
    primaryClass = "hep-ph",
    doi = "10.1088/1126-6708/2009/03/077",
    journal = "JHEP",
    volume = "03",
    pages = "077",
    year = "2009"
}

@article{Kamenik:2011vy,
    author = "Kamenik, Jernej F. and Smith, Christopher",
    title = "{FCNC portals to the dark sector}",
    eprint = "1111.6402",
    archivePrefix = "arXiv",
    primaryClass = "hep-ph",
    doi = "10.1007/JHEP03(2012)090",
    journal = "JHEP",
    volume = "03",
    pages = "090",
    year = "2012"
}

@article{LHCb:2021ykz,
    author = "Aaij, Roel and others",
    collaboration = "LHCb",
    title = "{Observation of the Mass Difference Between Neutral Charm-Meson Eigenstates}",
    eprint = "2106.03744",
    archivePrefix = "arXiv",
    primaryClass = "hep-ex",
    reportNumber = "LHCb-PAPER-2021-009, CERN-EP-2021-099",
    doi = "10.1103/PhysRevLett.127.111801",
    journal = "Phys. Rev. Lett.",
    volume = "127",
    number = "11",
    pages = "111801",
    year = "2021"
}

@article{KEDR:2018lhb,
    author = "Anashin, V. V. and others",
    collaboration = "KEDR",
    title = "{Measurement of $\Gamma_{ee}(J/\psi)$ with KEDR detector}",
    eprint = "1801.01958",
    archivePrefix = "arXiv",
    primaryClass = "hep-ex",
    doi = "10.1007/JHEP05(2018)119",
    journal = "JHEP",
    volume = "05",
    pages = "119",
    year = "2018",
    note = "[Addendum: JHEP 07, 112 (2020)]"
}

@article{Belle:2006tbq,
    author = "Satoyama, N. and others",
    collaboration = "Belle",
    title = "{A Search for the rare leptonic decays B+ ---\ensuremath{>} mu+ nu(mu) and B+ ---\ensuremath{>} e+ nu(nu)}",
    eprint = "hep-ex/0611045",
    archivePrefix = "arXiv",
    reportNumber = "BELLE-PREPRINT-2006-36, KEK-PREPRINT-2006-53",
    doi = "10.1016/j.physletb.2007.01.068",
    journal = "Phys. Lett. B",
    volume = "647",
    pages = "67--73",
    year = "2007"
}

@article{Belle:2010ouj,
    author = "Petric, M. and others",
    collaboration = "Belle",
    title = "{Search for leptonic decays of $D^0$ mesons}",
    eprint = "1003.2345",
    archivePrefix = "arXiv",
    primaryClass = "hep-ex",
    doi = "10.1103/PhysRevD.81.091102",
    journal = "Phys. Rev. D",
    volume = "81",
    pages = "091102",
    year = "2010"
}

@article{ALEPH:2005ab,
    author = "Schael, S. and others",
    collaboration = "ALEPH, DELPHI, L3, OPAL, SLD, LEP Electroweak Working Group, SLD Electroweak Group, SLD Heavy Flavour Group",
    title = "{Precision electroweak measurements on the $Z$ resonance}",
    eprint = "hep-ex/0509008",
    archivePrefix = "arXiv",
    reportNumber = "SLAC-R-774",
    doi = "10.1016/j.physrep.2005.12.006",
    journal = "Phys. Rept.",
    volume = "427",
    pages = "257--454",
    year = "2006"
}

@article{LHCb:2016gsk,
    author = "Aaij, Roel and others",
    collaboration = "LHCb",
    title = "{A precise measurement of the $B^0$ meson oscillation frequency}",
    eprint = "1604.03475",
    archivePrefix = "arXiv",
    primaryClass = "hep-ex",
    reportNumber = "LHCB-PAPER-2015-031, CERN-EP-2016-084",
    doi = "10.1140/epjc/s10052-016-4250-2",
    journal = "Eur. Phys. J. C",
    volume = "76",
    number = "7",
    pages = "412",
    year = "2016"
}

@article{Durieux:2014xla,
    author = "Durieux, Gauthier and Maltoni, Fabio and Zhang, Cen",
    title = "{Global approach to top-quark flavor-changing interactions}",
    eprint = "1412.7166",
    archivePrefix = "arXiv",
    primaryClass = "hep-ph",
    reportNumber = "CP3-14-85",
    doi = "10.1103/PhysRevD.91.074017",
    journal = "Phys. Rev. D",
    volume = "91",
    number = "7",
    pages = "074017",
    year = "2015"
}

@article{Narain:2022qud,
    author = "Narain, Meenakshi and others",
    title = "{The Future of US Particle Physics - The Snowmass 2021 Energy Frontier Report}",
    eprint = "2211.11084",
    archivePrefix = "arXiv",
    primaryClass = "hep-ex",
    reportNumber = "FERMILAB-FN-1219-PPD-T",
    month = "11",
    year = "2022"
}

@article{Alipour-Fard:2018rbc,
    author = "Alipour-Fard, Samuel and Craig, Nathaniel and Gori, Stefania and Koren, Seth and Redigolo, Diego",
    title = "{The second Higgs at the lifetime frontier}",
    eprint = "1812.09315",
    archivePrefix = "arXiv",
    primaryClass = "hep-ph",
    doi = "10.1007/JHEP07(2020)029",
    journal = "JHEP",
    volume = "07",
    pages = "029",
    year = "2020"
}

@article{Alipour-Fard:2018lsf,
    author = "Alipour-Fard, Samuel and Craig, Nathaniel and Jiang, Minyuan and Koren, Seth",
    title = "{Long Live the Higgs Factory: Higgs Decays to Long-Lived Particles at Future Lepton Colliders}",
    eprint = "1812.05588",
    archivePrefix = "arXiv",
    primaryClass = "hep-ph",
    doi = "10.1088/1674-1137/43/5/053101",
    journal = "Chin. Phys. C",
    volume = "43",
    number = "5",
    pages = "053101",
    year = "2019"
}

@article{Chiu:2021sgs,
    author = "Chiu, Wen Han and Liu, Zhen and Low, Matthew and Wang, Lian-Tao",
    title = "{Jet timing}",
    eprint = "2109.01682",
    archivePrefix = "arXiv",
    primaryClass = "hep-ph",
    reportNumber = "FERMILAB-PUB-21-372-T",
    doi = "10.1007/JHEP01(2022)014",
    journal = "JHEP",
    volume = "01",
    pages = "014",
    year = "2022"
}

@article{Liu:2020vur,
    author = "Liu, Jia and Liu, Zhen and Wang, Lian-Tao and Wang, Xiao-Ping",
    title = "{Enhancing Sensitivities to Long-lived Particles with High Granularity Calorimeters at the LHC}",
    eprint = "2005.10836",
    archivePrefix = "arXiv",
    primaryClass = "hep-ph",
    reportNumber = "EFI-20-8",
    doi = "10.1007/JHEP11(2020)066",
    journal = "JHEP",
    volume = "11",
    pages = "066",
    year = "2020"
}

@article{Liu:2018wte,
    author = "Liu, Jia and Liu, Zhen and Wang, Lian-Tao",
    title = "{Enhancing Long-Lived Particles Searches at the LHC with Precision Timing Information}",
    eprint = "1805.05957",
    archivePrefix = "arXiv",
    primaryClass = "hep-ph",
    reportNumber = "FERMILAB-PUB-18-173-T, EFI-18-7",
    doi = "10.1103/PhysRevLett.122.131801",
    journal = "Phys. Rev. Lett.",
    volume = "122",
    number = "13",
    pages = "131801",
    year = "2019"
}

@article{Blondel:2022qqo,
    author = "Blondel, A. and others",
    title = "{Searches for long-lived particles at the future FCC-ee}",
    eprint = "2203.05502",
    archivePrefix = "arXiv",
    primaryClass = "hep-ex",
    doi = "10.3389/fphy.2022.967881",
    journal = "Front. in Phys.",
    volume = "10",
    pages = "967881",
    year = "2022"
}

@article{Maltoni:2022bqs,
    author = "Maltoni, F. and others",
    title = "{TF07 Snowmass Report: Theory of Collider Phenomena}",
    eprint = "2210.02591",
    archivePrefix = "arXiv",
    primaryClass = "hep-ph",
    reportNumber = "FERMILAB-FN-1203-QIS",
    month = "10",
    year = "2022"
}

@article{Alimena:2019zri,
    author = "Alimena, Juliette and others",
    title = "{Searching for long-lived particles beyond the Standard Model at the Large Hadron Collider}",
    eprint = "1903.04497",
    archivePrefix = "arXiv",
    primaryClass = "hep-ex",
    doi = "10.1088/1361-6471/ab4574",
    journal = "J. Phys. G",
    volume = "47",
    number = "9",
    pages = "090501",
    year = "2020"
}

@article{Fox:2007in,
    author = "Fox, Patrick J. and Ligeti, Zoltan and Papucci, Michele and Perez, Gilad and Schwartz, Matthew D.",
    title = "{Deciphering top flavor violation at the LHC with $B$ factories}",
    eprint = "0704.1482",
    archivePrefix = "arXiv",
    primaryClass = "hep-ph",
    reportNumber = "UCB-PTH-07-06, YITP-SB-07-11",
    doi = "10.1103/PhysRevD.78.054008",
    journal = "Phys. Rev. D",
    volume = "78",
    pages = "054008",
    year = "2008"
}

@article{CMS:2012wao,
    author = "Chatrchyan, Serguei and others",
    collaboration = "CMS",
    title = "{Search for Flavor Changing Neutral Currents in Top Quark Decays in pp Collisions at 7 TeV}",
    eprint = "1208.0957",
    archivePrefix = "arXiv",
    primaryClass = "hep-ex",
    reportNumber = "CMS-TOP-11-028, CERN-PH-EP-2012-206",
    doi = "10.1016/j.physletb.2012.12.045",
    journal = "Phys. Lett. B",
    volume = "718",
    pages = "1252--1272",
    year = "2013"
}

@article{CMS:2013knb,
    author = "Chatrchyan, Serguei and others",
    collaboration = "CMS",
    title = "{Search for Flavor-Changing Neutral Currents in Top-Quark Decays $t \to Zq$ in $pp$ Collisions at $\sqrt{s}=8$  TeV}",
    eprint = "1312.4194",
    archivePrefix = "arXiv",
    primaryClass = "hep-ex",
    reportNumber = "CMS-TOP-12-037, CERN-PH-EP-2013-208",
    doi = "10.1103/PhysRevLett.112.171802",
    journal = "Phys. Rev. Lett.",
    volume = "112",
    number = "17",
    pages = "171802",
    year = "2014"
}

@article{CMS:2016obj,
    author = "Khachatryan, Vardan and others",
    collaboration = "CMS",
    title = "{Search for top quark decays via Higgs-boson-mediated flavor-changing neutral currents in pp collisions at $ \sqrt{s}=8 $ TeV}",
    eprint = "1610.04857",
    archivePrefix = "arXiv",
    primaryClass = "hep-ex",
    reportNumber = "CMS-TOP-13-017, CERN-EP-2016-208",
    doi = "10.1007/JHEP02(2017)079",
    journal = "JHEP",
    volume = "02",
    pages = "079",
    year = "2017"
}

@article{CMS:2017wcz,
    author = "Sirunyan, Albert M and others",
    collaboration = "CMS",
    title = "{Search for associated production of a Z boson with a single top quark and for tZ flavour-changing interactions in pp collisions at $ \sqrt{s}=8 $ TeV}",
    eprint = "1702.01404",
    archivePrefix = "arXiv",
    primaryClass = "hep-ex",
    reportNumber = "CMS-TOP-12-039, CERN-EP-2016-324",
    doi = "10.1007/JHEP07(2017)003",
    journal = "JHEP",
    volume = "07",
    pages = "003",
    year = "2017"
}

@article{CMS:2017bhz,
    author = "Sirunyan, Albert M and others",
    collaboration = "CMS",
    title = "{Search for the flavor-changing neutral current interactions of the top quark and the Higgs boson which decays into a pair of b quarks at $\sqrt{s}=$ 13 TeV}",
    eprint = "1712.02399",
    archivePrefix = "arXiv",
    primaryClass = "hep-ex",
    reportNumber = "CMS-TOP-17-003, CERN-EP-2017-309",
    doi = "10.1007/JHEP06(2018)102",
    journal = "JHEP",
    volume = "06",
    pages = "102",
    year = "2018"
}

@article{CMS:2021gfa,
    author = "Tumasyan, Armen and others",
    collaboration = "CMS",
    title = "{Search for flavor-changing neutral current interactions of the top quark and the Higgs boson decaying to a bottom quark-antiquark pair at $ \sqrt{s} $ = 13 TeV}",
    eprint = "2112.09734",
    archivePrefix = "arXiv",
    primaryClass = "hep-ex",
    reportNumber = "CMS-TOP-19-002, CERN-EP-2021-241",
    doi = "10.1007/JHEP02(2022)169",
    journal = "JHEP",
    volume = "02",
    pages = "169",
    year = "2022"
}

@article{ATLAS:2012hfh,
    author = "Aad, Georges and others",
    collaboration = "ATLAS",
    title = "{A search for flavour changing neutral currents in top-quark decays in $pp$ collision data collected with the ATLAS detector at $\sqrt{s}=7$ TeV}",
    eprint = "1206.0257",
    archivePrefix = "arXiv",
    primaryClass = "hep-ex",
    reportNumber = "CERN-PH-EP-2012-139, SLAC-PUB-16110",
    doi = "10.1007/JHEP09(2012)139",
    journal = "JHEP",
    volume = "09",
    pages = "139",
    year = "2012"
}

@article{ATLAS:2015vhj,
    author = "Aad, Georges and others",
    collaboration = "ATLAS",
    title = "{Search for flavour-changing neutral current top-quark decays to $qZ$ in $pp$ collision data collected with the ATLAS detector at $\sqrt s =8$  TeV}",
    eprint = "1508.05796",
    archivePrefix = "arXiv",
    primaryClass = "hep-ex",
    reportNumber = "CERN-PH-EP-2015-183",
    doi = "10.1140/epjc/s10052-015-3851-5",
    journal = "Eur. Phys. J. C",
    volume = "76",
    number = "1",
    pages = "12",
    year = "2016"
}

@article{Kong:2014jwa,
    author = "Kong, Kyoungchul and Lee, Hye-Sung and Park, Myeonghun",
    title = "{Dark decay of the top quark}",
    eprint = "1401.5020",
    archivePrefix = "arXiv",
    primaryClass = "hep-ph",
    reportNumber = "JLAB-THY-14-1840",
    doi = "10.1103/PhysRevD.89.074007",
    journal = "Phys. Rev. D",
    volume = "89",
    number = "7",
    pages = "074007",
    year = "2014"
}

@article{Kim:2014ana,
    author = "Kim, Doojin and Lee, Hye-Sung and Park, Myeonghun",
    title = "{Invisible dark gauge boson search in top decays using a kinematic method}",
    eprint = "1411.0668",
    archivePrefix = "arXiv",
    primaryClass = "hep-ph",
    reportNumber = "CERN-PH-TH-2014-209, APCTP-PRE2014-013, IPMU14-0330",
    doi = "10.1007/JHEP03(2015)134",
    journal = "JHEP",
    volume = "03",
    pages = "134",
    year = "2015"
}

@article{Alcaide:2019pnf,
    author = "Alcaide, Julien and Banerjee, Shankha and Chala, Mikael and Titov, Arsenii",
    title = "{Probes of the Standard Model effective field theory extended with a right-handed neutrino}",
    eprint = "1905.11375",
    archivePrefix = "arXiv",
    primaryClass = "hep-ph",
    reportNumber = "IPPP/19/43, FTUV-19-0524, IFIC/19-28",
    doi = "10.1007/JHEP08(2019)031",
    journal = "JHEP",
    volume = "08",
    pages = "031",
    year = "2019"
}

@article{Belle:2013hlo,
    author = "Sibidanov, A. and others",
    collaboration = "Belle",
    title = "{Study of Exclusive $B \to X_u \ell \nu$ Decays and Extraction of $\|V_{ub}\|$ using Full Reconstruction Tagging at the Belle Experiment}",
    eprint = "1306.2781",
    archivePrefix = "arXiv",
    primaryClass = "hep-ex",
    reportNumber = "BELLE-PREPRINT-2013-9, KEK-PREPRINT-2013-8",
    doi = "10.1103/PhysRevD.88.032005",
    journal = "Phys. Rev. D",
    volume = "88",
    number = "3",
    pages = "032005",
    year = "2013"
}

@article{ATLAS:2015ncl,
    author = "Aad, Georges and others",
    collaboration = "ATLAS",
    title = "{Search for flavour-changing neutral current top quark decays $t\to Hq$ in $pp$ collisions at $\sqrt{s}=8$ TeV with the ATLAS detector}",
    eprint = "1509.06047",
    archivePrefix = "arXiv",
    primaryClass = "hep-ex",
    reportNumber = "CERN-PH-EP-2015-229",
    doi = "10.1007/JHEP12(2015)061",
    journal = "JHEP",
    volume = "12",
    pages = "061",
    year = "2015"
}

@article{ATLAS:2017tas,
    author = "Aaboud, Morad and others",
    collaboration = "ATLAS",
    title = "{Search for top quark decays $t\rightarrow qH$, with $H\to\gamma\gamma$, in $\sqrt{s}=13$ TeV $pp$ collisions using the ATLAS detector}",
    eprint = "1707.01404",
    archivePrefix = "arXiv",
    primaryClass = "hep-ex",
    reportNumber = "CERN-EP-2017-118",
    doi = "10.1007/JHEP10(2017)129",
    journal = "JHEP",
    volume = "10",
    pages = "129",
    year = "2017"
}

@article{ATLAS:2018zsq,
    author = "Aaboud, M. and others",
    collaboration = "ATLAS",
    title = "{Search for flavour-changing neutral current top-quark decays $t\to qZ$ in proton-proton collisions at $\sqrt{s}=13$ TeV with the ATLAS detector}",
    eprint = "1803.09923",
    archivePrefix = "arXiv",
    primaryClass = "hep-ex",
    reportNumber = "CERN-EP-2018-018",
    doi = "10.1007/JHEP07(2018)176",
    journal = "JHEP",
    volume = "07",
    pages = "176",
    year = "2018"
}

@article{ATLAS:2018xxe,
    author = "Aaboud, Morad and others",
    collaboration = "ATLAS",
    title = "{Search for flavor-changing neutral currents in top quark decays $t\to Hc$ and $t \to Hu$ in multilepton final states in proton-proton collisions at $\sqrt{s}= 13$ TeV with the ATLAS detector}",
    eprint = "1805.03483",
    archivePrefix = "arXiv",
    primaryClass = "hep-ex",
    reportNumber = "CERN-EP-2018-067",
    doi = "10.1103/PhysRevD.98.032002",
    journal = "Phys. Rev. D",
    volume = "98",
    number = "3",
    pages = "032002",
    year = "2018"
}

@article{ATLAS:2018jqi,
    author = "Aaboud, Morad and others",
    collaboration = "ATLAS",
    title = "{Search for top-quark decays $t \to Hq$ with 36 fb$^{-1}$ of $pp$ collision data at $\sqrt{s}=13$ TeV with the ATLAS detector}",
    eprint = "1812.11568",
    archivePrefix = "arXiv",
    primaryClass = "hep-ex",
    reportNumber = "CERN-EP-2018-295",
    doi = "10.1007/JHEP05(2019)123",
    journal = "JHEP",
    volume = "05",
    pages = "123",
    year = "2019"
}

@article{ATLAS:2022per,
    collaboration = "ATLAS",
    title = "{Search for flavour-changing neutral-current couplings between the top quark and the photon with the ATLAS detector at $\sqrt{s} = 13$ TeV}",
    eprint = "2205.02537",
    archivePrefix = "arXiv",
    primaryClass = "hep-ex",
    reportNumber = "CERN-EP-2022-042",
    doi = "10.1016/j.physletb.2022.137379",
    month = "5",
    year = "2022"
}

@article{Barzinji:2018xvu,
    author = "Barzinji, Abdurrahman and Trott, Michael and Vasudevan, Anagha",
    title = "{Equations of Motion for the Standard Model Effective Field Theory: Theory and Applications}",
    eprint = "1806.06354",
    archivePrefix = "arXiv",
    primaryClass = "hep-ph",
    doi = "10.1103/PhysRevD.98.116005",
    journal = "Phys. Rev. D",
    volume = "98",
    number = "11",
    pages = "116005",
    year = "2018"
}

@article{ATLAS:2022vhr,
    collaboration = "ATLAS",
    title = "{Search for displaced photons produced in exotic decays of the Higgs boson using 13 TeV $pp$ collisions with the ATLAS detector}",
    eprint = "2209.01029",
    archivePrefix = "arXiv",
    primaryClass = "hep-ex",
    reportNumber = "CERN-EP-2022-096",
    month = "9",
    year = "2022"
}

@article{ATLAS:2022gzn,
    collaboration = "ATLAS",
    title = "{Search for flavour-changing neutral current interactions of the top quark and the Higgs boson in events with a pair of $\tau$-leptons in pp collisions at $\sqrt{s}=13$ TeV with the ATLAS detector}",
    eprint = "2208.11415",
    archivePrefix = "arXiv",
    primaryClass = "hep-ex",
    reportNumber = "CERN-EP-2022-123",
    month = "8",
    year = "2022"
}

@article{ATLAS:2023mcc,
    collaboration = "ATLAS",
    title = "{Search for a new scalar resonance in flavour-changing neutral-current top-quark decays $t \rightarrow qX$ ($q=u,c$), with $X \rightarrow b\bar{b}$, in proton-proton collisions at $\sqrt{s}=13$ TeV with the ATLAS detector}",
    eprint = "2301.03902",
    archivePrefix = "arXiv",
    primaryClass = "hep-ex",
    reportNumber = "CERN-EP-2022-176",
    month = "1",
    year = "2023"
}

@article{CMS:2022ztx,
    author = "Tumasyan, Armen and others",
    collaboration = "CMS",
    title = "{Search for charged-lepton flavor violation in top quark production and decay in pp collisions at $ \sqrt{s} $ = 13 TeV}",
    eprint = "2201.07859",
    archivePrefix = "arXiv",
    primaryClass = "hep-ex",
    reportNumber = "CMS-TOP-19-006, CERN-EP-2021-248",
    doi = "10.1007/JHEP06(2022)082",
    journal = "JHEP",
    volume = "06",
    pages = "082",
    year = "2022"
}

@article{Cremer:2023gne,
    author = {Cremer, Lucas and Erdmann, Johannes and Harnik, Roni and Sp\"ah, Jan Lukas and Stamou, Emmanuel},
    title = "{Leveraging on-shell interference to search for FCNCs of the top quark and the Z boson}",
    eprint = "2305.12172",
    archivePrefix = "arXiv",
    primaryClass = "hep-ph",
    reportNumber = "DO-TH 23/04",
    month = "5",
    year = "2023"
}

@article{Harlander:2020cyh,
    author = "Harlander, R. V. and Klein, S. Y. and Lipp, M.",
    title = "{FeynGame}",
    eprint = "2003.00896",
    archivePrefix = "arXiv",
    primaryClass = "physics.ed-ph",
    reportNumber = "TTK-20-04",
    doi = "10.1016/j.cpc.2020.107465",
    journal = "Comput. Phys. Commun.",
    volume = "256",
    pages = "107465",
    year = "2020"
}

@article{Kublbeck:1990xc,
    author = "Kublbeck, J. and Bohm, M. and Denner, Ansgar",
    title = "{Feyn Arts: Computer Algebraic Generation of Feynman Graphs and Amplitudes}",
    reportNumber = "Print-90-0144 (WURZBURG)",
    doi = "10.1016/0010-4655(90)90001-H",
    journal = "Comput. Phys. Commun.",
    volume = "60",
    pages = "165--180",
    year = "1990"
}

@article{Hahn:2000kx,
    author = "Hahn, Thomas",
    title = "{Generating Feynman diagrams and amplitudes with FeynArts 3}",
    eprint = "hep-ph/0012260",
    archivePrefix = "arXiv",
    reportNumber = "KA-TP-23-2000",
    doi = "10.1016/S0010-4655(01)00290-9",
    journal = "Comput. Phys. Commun.",
    volume = "140",
    pages = "418--431",
    year = "2001"
}

@article{Christensen:2008py,
    author = "Christensen, Neil D. and Duhr, Claude",
    title = "{FeynRules - Feynman rules made easy}",
    eprint = "0806.4194",
    archivePrefix = "arXiv",
    primaryClass = "hep-ph",
    reportNumber = "MSUHEP-080616, CP3-08-20",
    doi = "10.1016/j.cpc.2009.02.018",
    journal = "Comput. Phys. Commun.",
    volume = "180",
    pages = "1614--1641",
    year = "2009"
}

@article{Alloul:2013bka,
    author = "Alloul, Adam and Christensen, Neil D. and Degrande, C\'eline and Duhr, Claude and Fuks, Benjamin",
    title = "{FeynRules  2.0 - A complete toolbox for tree-level phenomenology}",
    eprint = "1310.1921",
    archivePrefix = "arXiv",
    primaryClass = "hep-ph",
    reportNumber = "CERN-PH-TH-2013-239, MCNET-13-14, IPPP-13-71, DCPT-13-142, PITT-PACC-1308",
    doi = "10.1016/j.cpc.2014.04.012",
    journal = "Comput. Phys. Commun.",
    volume = "185",
    pages = "2250--2300",
    year = "2014"
}

@article{Hahn:1998yk,
    author = "Hahn, T. and Perez-Victoria, M.",
    title = "{Automatized one loop calculations in four-dimensions and D-dimensions}",
    eprint = "hep-ph/9807565",
    archivePrefix = "arXiv",
    reportNumber = "UG-FT-87-98, KA-TP-7-1998",
    doi = "10.1016/S0010-4655(98)00173-8",
    journal = "Comput. Phys. Commun.",
    volume = "118",
    pages = "153--165",
    year = "1999"
}

@article{Alwall:2014hca,
    author = "Alwall, J. and Frederix, R. and Frixione, S. and Hirschi, V. and Maltoni, F. and Mattelaer, O. and Shao, H. -S. and Stelzer, T. and Torrielli, P. and Zaro, M.",
    title = "{The automated computation of tree-level and next-to-leading order differential cross sections, and their matching to parton shower simulations}",
    eprint = "1405.0301",
    archivePrefix = "arXiv",
    primaryClass = "hep-ph",
    reportNumber = "CERN-PH-TH-2014-064, CP3-14-18, LPN14-066, MCNET-14-09, ZU-TH-14-14",
    doi = "10.1007/JHEP07(2014)079",
    journal = "JHEP",
    volume = "07",
    pages = "079",
    year = "2014"
}

@article{Alwall:2014bza,
    author = {Alwall, Johan and Duhr, Claude and Fuks, Benjamin and Mattelaer, Olivier and \"Ozt\"urk, Deniz Gizem and Shen, Chia-Hsien},
    title = "{Computing decay rates for new physics theories with FeynRules  and MadGraph 5\_aMC@NLO}",
    eprint = "1402.1178",
    archivePrefix = "arXiv",
    primaryClass = "hep-ph",
    reportNumber = "CERN-PH-TH-2014-020, MCNET-14-03, IPPP-14-10, DCPT-14-20, CP3-14-11, CALT-68-2877, ZU-TH-02-14",
    doi = "10.1016/j.cpc.2015.08.031",
    journal = "Comput. Phys. Commun.",
    volume = "197",
    pages = "312--323",
    year = "2015"
}

@article{ATLAS:2018gfm,
    author = "Aaboud, Morad and others",
    collaboration = "ATLAS",
    title = "{Search for charged Higgs bosons decaying via $H^{\pm} \to \tau^{\pm}\nu_{\tau}$ in the $\tau$+jets and $\tau$+lepton final states with 36 fb$^{-1}$ of $pp$ collision data recorded at $\sqrt{s} = 13$ TeV with the ATLAS experiment}",
    eprint = "1807.07915",
    archivePrefix = "arXiv",
    primaryClass = "hep-ex",
    reportNumber = "CERN-EP-2018-148",
    doi = "10.1007/JHEP09(2018)139",
    journal = "JHEP",
    volume = "09",
    pages = "139",
    year = "2018"
}

@article{CMS:2022jqc,
    collaboration = "CMS",
    title = "{Search for a charged Higgs boson decaying into a heavy neutral Higgs boson and a W boson in proton-proton collisions at $\sqrt{s}$ = 13 TeV}",
    eprint = "2207.01046",
    archivePrefix = "arXiv",
    primaryClass = "hep-ex",
    reportNumber = "CMS-HIG-21-010, CERN-EP-2022-125",
    month = "7",
    year = "2022"
}

@article{CMS:2019bfg,
    author = "Sirunyan, Albert M and others",
    collaboration = "CMS",
    title = "{Search for charged Higgs bosons in the H$^{\pm}$ $\to$ $\tau^{\pm}\nu_\tau$ decay channel in proton-proton collisions at $\sqrt{s} =$ 13 TeV}",
    eprint = "1903.04560",
    archivePrefix = "arXiv",
    primaryClass = "hep-ex",
    reportNumber = "CMS-HIG-18-014, CERN-EP-2019-025",
    doi = "10.1007/JHEP07(2019)142",
    journal = "JHEP",
    volume = "07",
    pages = "142",
    year = "2019"
}

@article{ATLAS:2023bzb,
    collaboration = "ATLAS",
    title = "{Search for a light charged Higgs boson in $t \rightarrow H^{\pm}b$ decays, with $H^{\pm} \rightarrow cb$, in the lepton+jets final state in proton-proton collisions at $\sqrt{s}=13$ TeV with the ATLAS detector}",
    eprint = "2302.11739",
    archivePrefix = "arXiv",
    primaryClass = "hep-ex",
    reportNumber = "CERN-EP-2022-207",
    month = "2",
    year = "2023"
}

@article{Degrande:2011ua,
    author = "Degrande, Celine and Duhr, Claude and Fuks, Benjamin and Grellscheid, David and Mattelaer, Olivier and Reiter, Thomas",
    title = "{UFO - The Universal FeynRules Output}",
    eprint = "1108.2040",
    archivePrefix = "arXiv",
    primaryClass = "hep-ph",
    reportNumber = "CP3-11-25, IPHC-PHENO-11-04, IPPP-11-39, DCPT-11-78, MPP-2011-68",
    doi = "10.1016/j.cpc.2012.01.022",
    journal = "Comput. Phys. Commun.",
    volume = "183",
    pages = "1201--1214",
    year = "2012"
}

@article{Darme:2023jdn,
    author = "Darm\'e, Luc and others",
    title = "{UFO 2.0 -- The Universal Feynman Output format}",
    eprint = "2304.09883",
    archivePrefix = "arXiv",
    primaryClass = "hep-ph",
    reportNumber = "BONN-TH-2023-03, DESY-23-051, FERMILAB-PUB-23-138-T, KA-TP-06-2023,
  MCNET-23-06, P3H-23-023, TIF-UNIMI-2023-11",
    month = "4",
    year = "2023"
}

@article{Czakon:2011xx,
    author = "Czakon, Michal and Mitov, Alexander",
    title = "{Top++: A Program for the Calculation of the Top-Pair Cross-Section at Hadron Colliders}",
    eprint = "1112.5675",
    archivePrefix = "arXiv",
    primaryClass = "hep-ph",
    reportNumber = "CERN-PH-TH-2011-303, TTK-11-58",
    doi = "10.1016/j.cpc.2014.06.021",
    journal = "Comput. Phys. Commun.",
    volume = "185",
    pages = "2930",
    year = "2014"
}

@article{Froggatt:1978nt,
    author = "Froggatt, C. D. and Nielsen, Holger Bech",
    title = "{Hierarchy of Quark Masses, Cabibbo Angles and CP Violation}",
    reportNumber = "CERN-TH-2519",
    doi = "10.1016/0550-3213(79)90316-X",
    journal = "Nucl. Phys. B",
    volume = "147",
    pages = "277--298",
    year = "1979"
}

@article{Altmannshofer:2023bfk,
    author = "Altmannshofer, Wolfgang and Gori, Stefania and Lehmann, Benjamin V. and Zuo, Jianhong",
    title = "{UV physics from IR features: New prospects from top flavor violation}",
    eprint = "2303.00781",
    archivePrefix = "arXiv",
    primaryClass = "hep-ph",
    reportNumber = "MIT-CTP/5534",
    doi = "10.1103/PhysRevD.107.095025",
    journal = "Phys. Rev. D",
    volume = "107",
    number = "9",
    pages = "095025",
    year = "2023"
}

@article{Banerjee:2018fsx,
    author = "Banerjee, Shankha and Chala, Mikael and Spannowsky, Michael",
    title = "{Top quark FCNCs in extended Higgs sectors}",
    eprint = "1806.02836",
    archivePrefix = "arXiv",
    primaryClass = "hep-ph",
    reportNumber = "IPPP-18-41",
    doi = "10.1140/epjc/s10052-018-6150-0",
    journal = "Eur. Phys. J. C",
    volume = "78",
    number = "8",
    pages = "683",
    year = "2018"
}

@article{Bhattacharyya:2022ciw,
    author = "Bhattacharyya, Gautam and Dwivedi, Siddharth and Ghosh, Dilip Kumar and Saha, Gourab and Sarkar, Subir",
    title = "{Searching for exotic Higgs bosons at the LHC}",
    eprint = "2202.01068",
    archivePrefix = "arXiv",
    primaryClass = "hep-ph",
    doi = "10.1103/PhysRevD.106.055032",
    journal = "Phys. Rev. D",
    volume = "106",
    number = "5",
    pages = "055032",
    year = "2022"
}

@article{Bhattacharyya:2022umc,
    author = "Bhattacharyya, Gautam and Chakraborty, Indrani and Ghosh, Dilip Kumar and Jha, Tapoja and Saha, Gourab",
    title = "{Searching for exotic Higgs bosons from top quark decays at the HL-LHC}",
    eprint = "2212.09061",
    archivePrefix = "arXiv",
    primaryClass = "hep-ph",
    month = "12",
    year = "2022"
}

@article{Song:2023jqm,
    author = "Song, Huayang and Sun, Hao and Yu, Jiang-Hao",
    title = "{Complete EFT Operator Bases for Dark Matter and Weakly-Interacting Light Particle}",
    eprint = "2306.05999",
    archivePrefix = "arXiv",
    primaryClass = "hep-ph",
    month = "6",
    year = "2023"
}

@article{Aebischer:2018iyb,
    author = "Aebischer, Jason and Kumar, Jacky and Stangl, Peter and Straub, David M.",
    title = "{A Global Likelihood for Precision Constraints and Flavour Anomalies}",
    eprint = "1810.07698",
    archivePrefix = "arXiv",
    primaryClass = "hep-ph",
    doi = "10.1140/epjc/s10052-019-6977-z",
    journal = "Eur. Phys. J. C",
    volume = "79",
    number = "6",
    pages = "509",
    year = "2019"
}

@article{DeBlas:2019ehy,
    author = "De Blas, J. and others",
    title = "{$\texttt{HEPfit}$: a code for the combination of indirect and direct constraints on high energy physics models}",
    eprint = "1910.14012",
    archivePrefix = "arXiv",
    primaryClass = "hep-ph",
    reportNumber = "CERN-TH-2019-178, CPHT-RR060.102019, DESY-19-184, DESY 19-184, FTUV/19-1031, KEK-TH-2163, LPT-Orsay-19-36, PSI-PR-19-22, UCI-TR-2019-26, IFIC/19-44",
    doi = "10.1140/epjc/s10052-020-7904-z",
    journal = "Eur. Phys. J. C",
    volume = "80",
    number = "5",
    pages = "456",
    year = "2020"
}

@article{Straub:2018kue,
    author = "Straub, David M.",
    title = "{flavio: a Python package for flavour and precision phenomenology in the Standard Model and beyond}",
    eprint = "1810.08132",
    archivePrefix = "arXiv",
    primaryClass = "hep-ph",
    month = "10",
    year = "2018"
}

@article{GAMBITFlavourWorkgroup:2017dbx,
    author = "Bernlochner, Florian U. and others",
    collaboration = "GAMBIT Flavour Workgroup",
    title = "{FlavBit: A GAMBIT module for computing flavour observables and likelihoods}",
    eprint = "1705.07933",
    archivePrefix = "arXiv",
    primaryClass = "hep-ph",
    reportNumber = "NORDITA-2017-077, CERN-TH-2017-167, NORDITA 2017-077, gambit-code-2017",
    doi = "10.1140/epjc/s10052-017-5157-2",
    journal = "Eur. Phys. J. C",
    volume = "77",
    number = "11",
    pages = "786",
    year = "2017"
}

@article{EOSAuthors:2021xpv,
    author = "van Dyk, Danny and others",
    collaboration = "EOS Authors",
    title = "{EOS: a software for flavor physics phenomenology}",
    eprint = "2111.15428",
    archivePrefix = "arXiv",
    primaryClass = "hep-ph",
    reportNumber = "EOS-2021-04, TUM-HEP 1371/21, P3H-21-094, SI-HEP-2021-32",
    doi = "10.1140/epjc/s10052-022-10177-4",
    journal = "Eur. Phys. J. C",
    volume = "82",
    number = "6",
    pages = "569",
    year = "2022"
}

@article{Bahl:2021str,
    author = "Bahl, Henning and Stefaniak, Tim and Wittbrodt, Jonas",
    title = "{The forgotten channels: charged Higgs boson decays to a W$^{±}$ and a non-SM-like Higgs boson}",
    eprint = "2103.07484",
    archivePrefix = "arXiv",
    primaryClass = "hep-ph",
    reportNumber = "DESY-21-035, DESY 21-035, LU TP 21-09",
    doi = "10.1007/JHEP06(2021)183",
    journal = "JHEP",
    volume = "06",
    pages = "183",
    year = "2021"
}

@article{Akeroyd:2016ymd,
    author = "Akeroyd, A. G. and others",
    title = "{Prospects for charged Higgs searches at the LHC}",
    eprint = "1607.01320",
    archivePrefix = "arXiv",
    primaryClass = "hep-ph",
    reportNumber = "CERN-TH-2016-152",
    doi = "10.1140/epjc/s10052-017-4829-2",
    journal = "Eur. Phys. J. C",
    volume = "77",
    number = "5",
    pages = "276",
    year = "2017"
}

@article{Li:2023btx,
    author = "Li, Z. and Arhrib, A. and Benbrik, R. and Krab, M. and Manaut, B. and Moretti, S. and Wang, Y. and Yan, Q. S.",
    title = "{Discovering a light charged Higgs boson via $W^{\pm *}$ + 4$b$ final states at the LHC}",
    eprint = "2305.05788",
    archivePrefix = "arXiv",
    primaryClass = "hep-ph",
    month = "5",
    year = "2023"
}

@article{Sanyal:2023pfs,
    author = "Sanyal, Prasenjit and Wang, Daohan",
    title = "{Probing the electroweak $4b + \ell + {\rlap{\,/}{E}_T}$ final state in type I 2HDM at the LHC}",
    eprint = "2305.00659",
    archivePrefix = "arXiv",
    primaryClass = "hep-ph",
    month = "5",
    year = "2023"
}

@article{Fu:2023sng,
    author = "Fu, ChunHao and Gao, Jun",
    title = "{Constraint for a light charged Higgs boson and its neutral fellows from top quark pairs at the LHC}",
    eprint = "2304.07782",
    archivePrefix = "arXiv",
    primaryClass = "hep-ph",
    month = "4",
    year = "2023"
}

@article{Mondal:2023wib,
    author = "Mondal, Tanmoy and Moretti, Stefano and Munir, Shoaib and Sanyal, Prasenjit",
    title = "{Electroweak multi-Higgs production: A smoking gun for the Type-I 2HDM}",
    eprint = "2304.07719",
    archivePrefix = "arXiv",
    primaryClass = "hep-ph",
    month = "4",
    year = "2023"
}

@article{Kim:2023lxc,
    author = "Kim, Jinheung and Lee, Soojin and Sanyal, Prasenjit and Song, Jeonghyeon and Wang, Daohan",
    title = "{\ensuremath{\tau}$^{±}$\ensuremath{\nu}\ensuremath{\gamma}\ensuremath{\gamma} and \ensuremath{\ell^\pm\ell^\pm\gamma\gamma\slashed{E}_TX} to probe the fermiophobic Higgs boson with high cutoff scales}",
    eprint = "2302.05467",
    archivePrefix = "arXiv",
    primaryClass = "hep-ph",
    doi = "10.1007/JHEP04(2023)083",
    journal = "JHEP",
    volume = "04",
    pages = "083",
    year = "2023"
}

@article{Bhatia:2022ugu,
    author = "Bhatia, Disha and Desai, Nishita and Dwivedi, Siddharth",
    title = "{Discovery prospects of a light charged Higgs near the fermiophobic region of Type-I 2HDM}",
    eprint = "2212.14363",
    archivePrefix = "arXiv",
    primaryClass = "hep-ph",
    doi = "10.1007/JHEP06(2023)100",
    journal = "JHEP",
    volume = "06",
    pages = "100",
    year = "2023"
}

@article{Krab:2022lih,
    author = "Krab, M. and Ouchemhou, M. and Arhrib, A. and Benbrik, R. and Manaut, B. and Yan, Qi-Shu",
    title = "{Single charged Higgs boson production at the LHC}",
    eprint = "2210.09416",
    archivePrefix = "arXiv",
    primaryClass = "hep-ph",
    doi = "10.1016/j.physletb.2023.137705",
    journal = "Phys. Lett. B",
    volume = "839",
    pages = "137705",
    year = "2023"
}

@article{Hu:2022gwd,
    author = "Hu, YaLu and Fu, ChunHao and Gao, Jun",
    title = "{Signature of a light charged Higgs boson from top quark pairs at the LHC}",
    eprint = "2206.05748",
    archivePrefix = "arXiv",
    primaryClass = "hep-ph",
    doi = "10.1103/PhysRevD.106.L071701",
    journal = "Phys. Rev. D",
    volume = "106",
    number = "7",
    pages = "L071701",
    year = "2022"
}

@article{Cheung:2022ndq,
    author = "Cheung, Kingman and Jueid, Adil and Kim, Jinheung and Lee, Soojin and Lu, Chih-Ting and Song, Jeonghyeon",
    title = "{Comprehensive study of the light charged Higgs boson in the type-I two-Higgs-doublet model}",
    eprint = "2201.06890",
    archivePrefix = "arXiv",
    primaryClass = "hep-ph",
    reportNumber = "KIAS-Q22002",
    doi = "10.1103/PhysRevD.105.095044",
    journal = "Phys. Rev. D",
    volume = "105",
    number = "9",
    pages = "095044",
    year = "2022"
}

@article{Balaji:2020qjg,
    author = "Balaji, Shyam",
    title = "{$CP$ asymmetries in the rare top decays $t\to c\gamma$ and $t\to c g$}",
    eprint = "2009.03315",
    archivePrefix = "arXiv",
    primaryClass = "hep-ph",
    doi = "10.1103/PhysRevD.102.113010",
    journal = "Phys. Rev. D",
    volume = "102",
    number = "11",
    pages = "113010",
    year = "2020"
}

@article{Mahlon:1994us,
    author = "Mahlon, Gregory and Parke, Stephen J.",
    title = "{Finite width effects in top quark decays}",
    eprint = "hep-ph/9412250",
    archivePrefix = "arXiv",
    reportNumber = "FERMILAB-PUB-94-406-T",
    doi = "10.1016/0370-2693(95)00083-W",
    journal = "Phys. Lett. B",
    volume = "347",
    pages = "394--398",
    year = "1995"
}

@article{Han:2013sea,
    author = "Han, Tao and Ruiz, Richard",
    title = "{Higgs Bosons from Top Quark Decays}",
    eprint = "1312.3324",
    archivePrefix = "arXiv",
    primaryClass = "hep-ph",
    reportNumber = "PITT-PACC-1313",
    doi = "10.1103/PhysRevD.89.074045",
    journal = "Phys. Rev. D",
    volume = "89",
    number = "7",
    pages = "074045",
    year = "2014"
}

@article{Dong:2011rh,
    author = "Dong, Zhe and Durieux, Gauthier and Gerard, Jean-Marc and Han, Tao and Maltoni, Fabio",
    title = "{Baryon number violation at the LHC: the top option}",
    eprint = "1107.3805",
    archivePrefix = "arXiv",
    primaryClass = "hep-ph",
    reportNumber = "CP3-11-24, MADPH-11-1573",
    doi = "10.1103/PhysRevD.85.016006",
    journal = "Phys. Rev. D",
    volume = "85",
    pages = "016006",
    year = "2012"
}

@article{Andrea:2011ws,
    author = "Andrea, J. and Fuks, B. and Maltoni, F.",
    title = "{Monotops at the LHC}",
    eprint = "1106.6199",
    archivePrefix = "arXiv",
    primaryClass = "hep-ph",
    reportNumber = "CP3-11-21, IPHC-PHENO-11-03",
    doi = "10.1103/PhysRevD.84.074025",
    journal = "Phys. Rev. D",
    volume = "84",
    pages = "074025",
    year = "2011"
}

@article{Agram:2013koa,
    author = "Agram, Jean-Laurent and Andrea, Jeremy and Conte, Eric and Fuks, Benjamin and Gel\'e, Denis and Lansonneur, Pierre",
    title = "{Probing top anomalous couplings at the LHC with trilepton signatures in the single top mode}",
    eprint = "1304.5551",
    archivePrefix = "arXiv",
    primaryClass = "hep-ph",
    reportNumber = "CERN-PH-TH-2013-078, IPHC-PHENO-13-03",
    doi = "10.1016/j.physletb.2013.06.052",
    journal = "Phys. Lett. B",
    volume = "725",
    pages = "123--126",
    year = "2013"
}

@article{Mangano:2016jyj,
    author = "Mangano, M. L. and others",
    title = "{Physics at a 100 TeV pp Collider: Standard Model Processes}",
    eprint = "1607.01831",
    archivePrefix = "arXiv",
    primaryClass = "hep-ph",
    reportNumber = "CERN-TH-2016-112",
    doi = "10.23731/CYRM-2017-003.1",
    month = "7",
    year = "2016"
}

\end{document}